\documentclass{iopart}
\usepackage{iopams,amsfonts,amssymb,amsthm} 
\usepackage{geometry}                
\usepackage{graphicx}
\usepackage{rawfonts}
\usepackage{amssymb}
\usepackage{epstopdf}
\usepackage[usenames,dvipsnames,svgnames]{xcolor}
\input prepictex.tex
\input pictex.tex
\input postpictex.tex

\newtheorem{theo}{Theorem}
\newtheorem{lemm}{Lemma}

\def\Pr{\noindent \emph{Proof: }}
\def\qed{$\Box$}

\def\nor{\normalsize}

\def\sfrac#1#2{\hbox{\nor $\frac{#1}{#2}$}}
\def\Sfrac#1#2{\hbox{\large $\frac{#1}{#2}$}}

\def\Ref#1{(\ref{#1})}

\def\L{\left(} \def\R{\right)}
\def\LC{\left\{} \def\RC{\right\}}

\def\thin{\hspace{0.1mm}}

\begin{document}

\title{Force-induced desorption of uniform branched polymers}
\author{E J Janse van Rensburg$^*$ and S G  Whittington$^\dagger$ }
\address{
{}$^*$Department of Mathematics \& Statistics, York University, M3J 1P3, Toronto, Canada \\
{}$^\dagger$Department of Chemistry, University of Toronto, M5S 3H6, Toronto, Canada 
}

\begin{abstract}
We analyze the phase diagrams of self-avoiding walk models of uniform branched polymers
adsorbed at a surface and subject to an externally applied vertical pulling force which, at
critical values, desorbs the polymer.   In particular, models of adsorbed branched polymers with 
homeomorphism types stars,  tadpoles, dumbbells and combs are examined.  These models generalize earlier results on linear, ring and $3$-star polymers.  In the case of star
polymers we confirm a phase diagram with four phases (a free, an adsorbed, 
a ballistic, and a mixed
phase) first seen in reference \cite{Rensburg2018} for $3$-star polymers.  
The phase diagram of tadpoles may include
four phases (including a mixed phase) if the tadpole is pulled from the 
adsorbing surface by the end vertex
of its tail.  If it is instead pulled from the middle vertex of its head, then there are only three phases
(the mixed phase is absent).  For a dumbbell pulled from the middle vertex of a ring, there
are only three phases.  For combs with $t$ teeth there are four phases, independent of 
the value of $t$ for all $t \ge 1$.
\end{abstract}

\maketitle

\section{Introduction}
\label{sec:Introduction}
With the availability of micro-manipulation techniques such as atomic force microscopy that
allow single polymer molecules to be pulled  \cite{Haupt1999,Zhang2003}, 
there has been considerable interest
in developing theories of how polymers respond to applied forces
\cite{Beaton2015,GuttmannLawler,IoffeVelenik,IoffeVelenik2010,Rensburg2016a}.  
A case that has attracted a lot of attention is when the polymer is adsorbed at a surface and is 
desorbed by the action of the force \cite{Rensburg2013,Krawczyk2005,Krawczyk2004,Mishra2005}.
For a recent review see reference \cite{Orlandini}.  Several different models have been investigated
\cite{Orlandini} but we shall concentrate here on self-avoiding walks \cite{MadrasSlade} and
related systems.  For other related work, see \cite{Skvortsov2009,Binder2012}.
We give a brief review of the results for linear and ring polymers in  
Section \ref{sec:review}.  

Does it matter where the force is applied?  It turns out that sometimes it does
\cite{Rensburg2017} and sometimes it doesn't \cite{Rensburg2016b}.
It is also natural to ask if the results depend on the architecture of the polymer.  That is, do
ring polymers, or star polymers, etc, behave differently to linear polymers?  Ring 
polymers have been investigated in both two \cite{Beaton2017, Guttmann2018} and 
three dimensions \cite{Guttmann2018}.  The behaviour in three dimensions is qualitatively 
similar to that of linear polymers but, in two dimensions, there is an 
additional phase.  3-star polymers have been investigated in three dimensions 
\cite{Rensburg2018} and they also show a similar additional phase.

\begin{figure}[b]
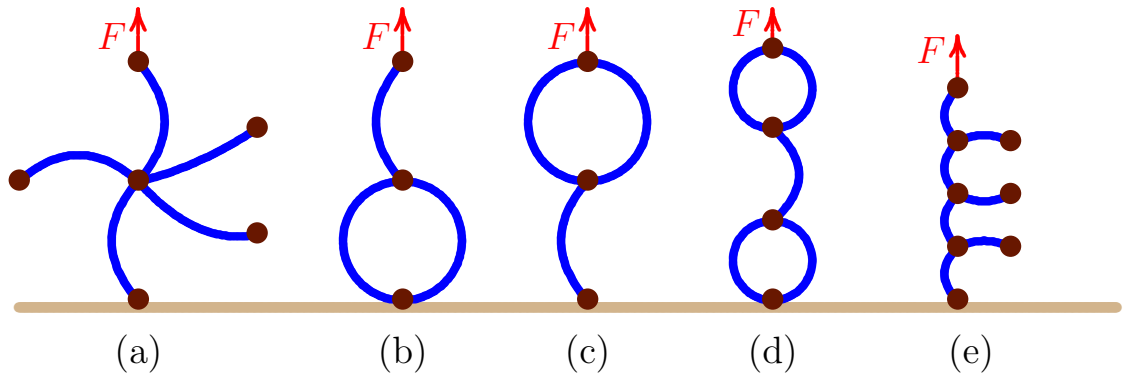

\beginpicture
\setcoordinatesystem units <1.0pt,1.0pt>
\setplotarea x from -140 to 150, y from -20 to 110
\setplotarea x from -100 to 250, y from 0 to 90

\color{Tan}
\setplotsymbol ({$\bullet$})
\plot -145 -3 270 -3 /

\setquadratic

\setplotsymbol ({\scriptsize$\bullet$})
\color{Blue}
\plot -100 0 -110 22.5 -100 45 /
\plot -100 45 -90 67.5 -100 90 /
\plot -100 45 -77.5 27.5 -55 25 /
\plot -100 45 -77.5 52.5 -55 65 /
\plot -100 45 -122.5 55 -145 45 /
\color{red}
\setplotsymbol ({\large$\cdot$})
\arrow <8pt> [.2,.67] from -100 90 to -100 110 
\color{Sepia}
\multiput {\huge$\bullet$} at -100 0 -100 45 -100 90 -55 25 -55 65 -145 45 /

\setplotsymbol ({\scriptsize$\bullet$})
\color{Blue}
\circulararc 360 degrees from 0 0 center at 0 22.5
\plot 0 45 -10 67.5 0 90 /
\color{red}
\setplotsymbol ({\large$\cdot$})
\arrow <8pt> [.2,.67] from 0 90 to 0 110
\color{Sepia}
\multiput {\huge$\bullet$} at 0 0 0 90 0 45 /

\setplotsymbol ({\scriptsize$\bullet$})
\color{Blue}
\circulararc 360 degrees from 70 45 center at 70 67.5
\plot 70 0 60 22.5 70 45 /
\color{red}
\setplotsymbol ({\large$\cdot$})
\arrow <8pt> [.2,.67] from 70 90 to 70 110
\color{Sepia}
\multiput {\huge$\bullet$} at 70 0 70 90 70 45 /

\setplotsymbol ({\scriptsize$\bullet$})
\color{Blue}
\circulararc 360 degrees from 140 0 center at 140 15
\circulararc 360 degrees from 140 65 center at 140 80
\plot 140 30 150 47.5 140 65  /
\color{red}
\setplotsymbol ({\large$\cdot$})
\arrow <8pt> [.2,.67] from 140 95 to 140 110
\color{Sepia}
\multiput {\huge$\bullet$} at 140 0 140 30 140 65 140 95 /

\setplotsymbol ({\scriptsize$\bullet$})
\color{Blue}
\plot 210 0 205 10 210 20   /
\plot 210 20 205 30 210 40   /
\plot 210 40 205 50 210 60   /
\plot 210 60 205 70 210 80   /

\plot 210 20 220 22.5 230 20 /
\plot 210 40 220 37.5 230 40 /
\plot 210 60 220 62.5 230 60 /
\color{red}
\setplotsymbol ({\large$\cdot$})
\arrow <8pt> [.2,.67] from 210 80 to 210 100
\color{Sepia}
\multiput {\huge$\bullet$} at 210 0 210 20 210 40 210 60 210 80
230 20 230 40 230 60   /

\color{red}
\multiput {\Large$F$} at -110 100 -10 100 60 100 130 105 200 95 /

\color{black}
\put {\Large(a)} at -100 -20 
\put {\Large(b)} at 0 -20 
\put {\Large(c)} at 70 -20 
\put {\Large(d)} at 140 -20 
\put {\Large(e)} at 215 -20 
\normalcolor
\setlinear
\endpicture
\caption{Models of uniform branched polymers pulled by a vertical force $F$: 
(a) an $f$-star pulled from an end-vertex, (b) a tadpole pulled from its end-vertex, 
(c) a tadpole pulled from the middle vertex of its ring, (d) a dumbbell pulled from the
middle vertex of a ring, (e) a comb pulled from the end of its backbone.}
\label{fig:models}  
\end{figure}

In this paper we examine a variety of different polymer architectures and also
investigate the effect of applying the force in different ways (see figure \ref{fig:models}).  
We identify the various phases in the force-temperature plane and investigate the conditions under
which additional phases are present.  In particular we look at various star polymers
(with different numbers of branches), extending the results in \cite{Rensburg2018},
pulled either at the central vertex or at a unit degree vertex.  We compare with several other homeomorphism types, including tadpoles and combs.

\section{A brief review}
\label{sec:review}

In this section we give a brief account of previous results, concentrating
on self-avoiding walks.  We shall need some of these results in the following sections.
We focus on the simple cubic lattice $\mathbb{Z}^3$ but the results for self-avoiding walks 
can be extended to $\mathbb{Z}^d$ for all $d \ge 2$.

Consider self-avoiding walks on $\mathbb{Z}^3$ where we attach the obvious
coordinate system $(x_1,x_2,x_3)$ so that lattice vertices have integer coordinates.
For an $n$-edge self-avoiding walk number the vertices $k=0,1, \ldots, n$ and 
write $x_i(k)$ for the $i$th coordinate of the $k$th vertex.  Write $c_n$ for the number of 
$n$-edge self-avoiding walks starting at the origin.  Hammersley 
\cite{Hammersley1957} showed 
the existence of the limit
\begin{equation}
\lim_{n\to\infty} \Sfrac{1}{n}  \log c_n = \inf_{n>0} \Sfrac{1}{n}  \log c_n \equiv \log \mu_3
\end{equation}
where the \emph{growth constant} $\mu_3$ satisfies the inequalities $3 < \mu_3 < 5$.  
Suppose that the self-avoiding walk satisfies the additional constraint that $x_3(k) \ge 0$ for $0 \le k \le n$.
We call these \emph{positive walks} (see figure \ref{fig:poswalk}) 
and write $c_n^+$ for the number of positive
walks with $n$ edges.  Then $\lim_{n\to\infty} \sfrac{1}{n} \log c_n^+ = \log \mu_3$
\cite{Whittington1975}.

Let $c_n^+(v,h)$ be the number of $n$-edge positive walks, starting at the origin,
having $v+1$ vertices in the plane $x_3=0$ and with $x_3(n)=h$.  We say that the 
walk has $v$ \emph{visits} and that the \emph{height} of the last vertex is $h$.
Define the partition function $C_n^+(a,y) = \sum_{v,h} c_n^+(v,h) a^vy^h$.
The limit 
\begin{equation}
\lim_{n\to\infty} \Sfrac{1}{n}  \log C_n^+(a,y) \equiv \psi(a,y)
\end{equation}
exists for all $a$ and $y$ \cite{Rensburg2013}.  $\psi(a,y)$ is the \emph{free energy} of the model.  We can write
$a=\exp[-\epsilon/k_BT]$ and $y=\exp[F/k_BT]$ where $\epsilon$ is the energy associated with
a vertex in the surface, $k_B$ is Boltzmann's constant, $T$ is the absolute temperature and
$F$ is the force normal to the surface (measured in energy units).

If we turn off the force by setting $y=1$ we have the pure adsorption problem and we 
write $\psi(a,1) = \kappa(a)$.  Then $\kappa(a)$ is a convex function of $\log a$ and 
therefore continuous.  There exists $a_c > 1$ such that $\kappa(a) =
\log \mu_3$ when $a \le a_c$ and $\kappa(a) > \log \mu_3$ when $a > a_c$ 
\cite{HTW,Rensburg1998,Madras}.
If we set $a=1$ so that the interaction energy with the surface is zero we can write
$\psi(1,y) = \lambda(y)$.  $\lambda(y)$ is a convex function of $\log y$ \cite{Rensburg2009},
$\lambda(y) = \log \mu_3$ for $ a \le 1$ and $\lambda(y) > \log \mu_3$ for $a > 1$
\cite{Beaton2015,IoffeVelenik}.

\begin{figure}[b]
\beginpicture
\setcoordinatesystem units <1.0pt,1.0pt>
\setplotarea x from -150 to 150, y from -20 to 110
\setplotarea x from -100 to 220, y from 0 to 90

\color{Tan}
\setplotsymbol ({$\bullet$})
\plot -145 -3 270 -3 /

\setplotsymbol ({\scriptsize$\bullet$})
\color{Blue}
\plot -100 0 -100 10 -90 10 -90 0 -80 0 -80 10 -80 20 -80 30 -90 30 -100 30 -110 30 
-120 30 -120 20 -110 20 -110 10 -110 0 -120 0 -130 0 -130 10 -130 20 -130 30 
-130 40 -130 50 -120 50 -110 50 -110 40 -100 40 -100 50 -90 50  -80 50 -80 40 -70 40 
-60 40 -60 50 -60 60 -70 60 -80 60 -90 60 -100 60 -110 60 -110 70 -100 70 -100 80 -110 80 
-110 90 /
\color{red}
\setplotsymbol ({\large$\cdot$})
\arrow <8pt> [.2,.67] from -110 90 to -110 110 
\color{Sepia}
\multiput {\huge$\bullet$} at -100 0 -110 90  /

\setplotsymbol ({\scriptsize$\bullet$})
\color{Blue}
\plot -20 70 -10 70 -10 60 0 60 10 60 20 60 20 50 10 50  0 50 0 40 0 30 10 30 10 40 20 40 
20 30  20 20 30 20 40 20 50 20 60 20 60 10 50 10 40 10 30 10 20 10 10 10 10 20 0 20 0 10 
0 0 10 0 20 0 30 0 40 0 50 0 60 0 70 0 70 10 80 10 80 0  /
\color{red}
\setplotsymbol ({\large$\cdot$})
\arrow <8pt> [.2,.67] from -20 70 to -20 100
\color{Sepia}
\multiput {\huge$\bullet$} at -20 70 80 0 /
\setdashes <3pt> \plot -20 0 -20 70 /
\setsolid

\setplotsymbol ({\scriptsize$\bullet$})
\color{Blue}
\plot 140 0 150 0 150 10 160 10 170 10 170 20 170 30 160 30 150 30 150 40 160 40 
160 50 170 50  170 40 180 40 180 50 180 60 180 70 170 70 170 80 180 80 190 80
200 80 200 70 200 60 190 60 190 50 200 50 210 50 210 40 210 30 200 30 190 30 180 30
180 20 190 20 200 20 200 10 200 0 210 0 210 10 220 10 220 0   /
\color{red}
\setplotsymbol ({\large$\cdot$})
\color{Sepia}
\multiput {\huge$\bullet$} at  140 0 220 0  /
\setdashes <3pt> \plot 140 0 140 80 /

\color{red}
\multiput {\Large$F$} at -120 100 -30  80   /

\color{black}
\put {\Large(a)} at -100 -20 
\put {\Large(b)} at 30 -20 
\put {\Large(c)} at 180 -20 
\normalcolor
\setlinear
\endpicture
\caption{Positive walks, bridges and loops.
(a) A positive walk pulled from its endpoint by a force $F$.
(b) A doubly unfolded bridge pulled from its endpoint by a force $F$. 
(c) An unfolded loop.}
\label{fig:poswalk}  
\end{figure}

Returning to the full problem for all values of $a$ and $y$ \cite{Rensburg2013}
\begin{equation}
\psi(a,y) = \max[\kappa(a),\lambda(y)]
\end{equation}
so $\psi(a,y) = \log \mu_3$ when $a \le a_c$ and $y \le 1$.  This is the 
\emph{free phase}.  There are phase boundaries
in the $(a,y)$-plane at $a=a_c$ for $y \le 1$, at $y=1$ for $a \le a_c$ and at
the solution of $\kappa(a) = \lambda(y) $ for $a \ge a_c$ and $y \ge 1$.  The phase diagram 
has three phases and the phase transition for $y > 1$ and $a > a_c$ between the \emph{adsorbed
phase} (where the free energy is $\kappa(a)$) and the \emph{ballistic phase} (where the 
free energy is $\lambda(y)$) is first order \cite{Guttmann2014}.

The corresponding problem for polygons in $\mathbb{Z}^3$ has been investigated
\cite{Guttmann2018}.  If we write $\psi^0(a,y)$ for the free energy then
\begin{equation}
\psi^0(a,y) = \max[\kappa(a),\lambda(\sqrt{y})]
\end{equation}
so the phase diagram is qualitatively similar though the phase boundary between the 
adsorbed and ballistic phases is in a different location.  In two dimensions there is 
evidence of a fourth phase \cite{Beaton2017,Guttmann2018} where the free energy depends
on both $a$ and $y$.

One class of branched polymers has been studied \cite{Rensburg2018}.  These are
uniform 3-stars with one vertex of degree 1 at the origin and pulled from another vertex of degree 1.
The free energy  $\sigma^{(3)}(a,y)$ is given by
\begin{equation}
\sigma^{(3)}(a,y) = \max\left\{\kappa(a), \Sfrac{1}{3}( 2\lambda(y) + \log \mu_3 ),
\Sfrac{1}{3} ( 2\kappa(a) + \lambda(y) )\right\}.
\end{equation}
The phase diagram has four phases, a free phase where the free energy is
$\log \mu_3$, an adsorbed phase where the free energy is $\kappa(a)$,
a ballistic phase where the free energy depends only on $y$ and a mixed phase where
the free energy depends on both $a$ and $y$.

In this paper we consider pulled adsorbing branched uniform networks in the half-lattice
$x_3\geq 0$.  One of the branches is rooted at the origin, and this will be the
\textit{attached branch} or \textit{attached arm} of the network.  The network is 
adsorbing in the plane $x_3=0$ which is the \textit{adsorbing plane}, and pulled
at another vertex by a vertical force $F$ in the $x_3$ direction.  In the notation above,
$y=e^{F/k_BT}$, and if $y<1$ then $F<0$ and the force is pushing the vertex towards
the adsorbing plane.  If $y>0$ then $F$ is a force pulling the vertex away from the adsorbing plane..

\section{Uniform stars}
\label{sec:stars}
We first introduce some notation and recall some results about stars 
that we shall need.  An \emph{$f$-star} is an embedding in a lattice
(usually in $\mathbb{Z}^3$) of a connected graph with no cycles,
with one vertex of degree $f$, $f$ vertices of degree 1, and all other vertices of 
degree 2.  Each of the sets of edges from the vertex of degree $f$ to a vertex of degree
1 is a \emph{branch} or \emph{arm} and we use the two terms
interchangeably.  If all the branches have the same number of edges the star
is a \emph{uniform star}.  We shall be concerned almost exclusively with 
uniform stars and we shall often omit the word uniform.  

We shall count embeddings in the cubic lattice, $\mathbb{Z}^3$,
of uniform stars modulo translation.
Write $s_n^{(f)}$ for the number of such embeddings 
with a total of $n$ edges.  Note that $f$ must divide $n$.  For the  
cubic lattice $\mathbb{Z}^3$ with $f = 3, \ldots, 6$, we know that 
\cite{WhittingtonSoteros1992}
\begin{equation}
\lim_{n\to\infty} \sfrac{1}{n} \log s_n^{(f)} = \log \mu_3
\label{eqn:stargrowth}
\end{equation}
where the limit $n\to\infty$ is taken through $n=fm$
(multiples of $f$ in $\mathbb{N}$)).  

\begin{figure}[t]
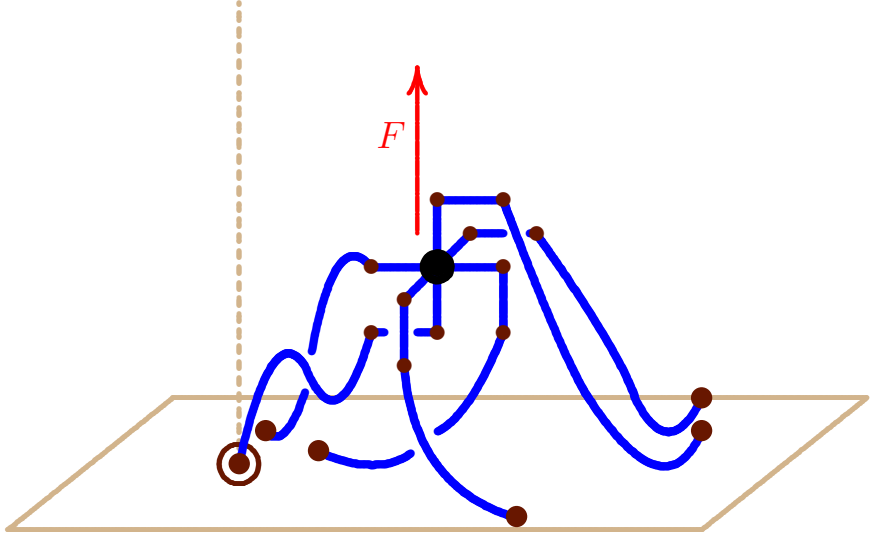

\beginpicture
\setcoordinatesystem units <2.5pt,2.5pt>
\setplotarea x from -85 to 100, y from -15 to 70
\setplotarea x from -65 to 60, y from 0 to 70

\color{Tan}
\setplotsymbol ({\LARGE$\cdot$})

\plot -65 -10 40 -10 65 10 -40 10 -65 -10 /

\setdashes <3pt>
\plot -30 0 -30 70  /
\setsolid
\setlinear

\color{Sepia}
\circulararc 360 degrees from -30 3 center at -30 0 

\setplotsymbol ({\scriptsize$\bullet$})
\color{Blue}

\plot -10 20 -8 20 / \plot -3 20  0 20 0 30  /
\plot 0 30 -10 30 /
\plot 0 30 10 30 10 20 /
\plot  0 30 -5 25 -5 15 /
\plot 0 30 5 35  10 35 / \plot 14 35 15 35 /
\plot 0 30 0 40 10 40 /

\setquadratic
\plot -30 0 -25 15  -20 15 -15 10 -10 20 /
\plot -10 30 -15 30 -19 17 / \plot -20 11  -23 5  -26 5  /
\plot -5 15  0 0 12 -8 /
\plot 10 20 5 10 0 5 /  \plot -4 1.75 -10  0  -18 2 /
\plot 10 40 18 20 23 10 33 0 40 5 /

\plot 15 35 25 20 30 10 35 5 40 10  / 

\color{Sepia}
\multiput {\Large$\bullet$} at -10 30 -10 20 0 20 0 40 -5 25 5 35 -5 15 10 30 15 35 10 40 10 20 /

\setplotsymbol ({\large$\cdot$})
\color{Sepia}
\multiput {\huge$\bullet$} at -30 0 -26 5  12 -8 -18 2   40 5 40 10 /
\color{Black}
\multiput {\huge$\bullet$} at 0 30  /
\setplotsymbol ({\scriptsize$\bullet$})
\circulararc 360 degrees from 0 32 center at 0 30 
\setplotsymbol ({\Large$\cdot$})
\color{red}
\arrow <10pt> [.2,.67]  from -3 35 to -3 60
\put {\Large$F$} at -7  50
\setplotsymbol ({\large$\cdot$})
\color{Tan}

\color{black}
\normalcolor
\endpicture
\caption{A schematic diagram of an adsorbing pulled $6$-star in the half cubic lattice.  
One branch (or arm) is fixed at the origin as
denoted, and the star is pulled by a force $F$ at its central node.  Vertices in the branches adsorb
in the adsorbing plane with activity $a$. Since each arm of this star has a visit in the adsorbing plane,
and it is pulled in its central node, this is a pulled $f$-spider. }
\label{figure1} 
 \end{figure}

We are primarily concerned with $f$-stars with a vertex of degree $1$
in the adsorbing plane $x_3=0$ (fixed at the origin).   The adsorbing
plane divides the lattice and the star is then confined to the upper half
lattice with an end vertex of one arm at the origin, and 
where it is pulled at another vertex by a vertical force $F$
(in the $x_3$-direction).  If the star is pulled at its central node \textit{and}
each arm has at least one visit in the adsorbing plane, then we call it
an $f$-\textit{spider} (and the arms are called \textit{legs}) -- see
figure \ref{figure1}.  Normally, the arms of the star are not constrained 
to have visits in the adsorbing plane,  and it is pulled either at its 
central node, or at another vertex of degree $1$.  We
shall consider all these cases below, namely pulled spiders, and stars
pulled at a central node, or at a vertex of degree $1$.

If the star has $v+1$ vertices in $x_3=0$ (these are \emph{visits} in the
adsorbing plane), then it is weighted by a factor $a^v$.  The \emph{height}
of the vertex where the star is pulled by a force $F$ is denoted by $h$ and 
the star will be weighted by a factor $y^h$.   We shall write $u_n^{(f)}(v,h)$
for the number of $f$-spiders with these conditions.  The partition function
of $f$-spiders is
\begin{equation}
U_n^{(f)}(a,y) = \sum_{v,h} u_n^{(f)}(v,h) \, a^v y^h.
\label{eqn:spiderPF}
\end{equation}

We shall write $s_n^{(f)}(v,h)$ for the number of $f$-stars pulled at a vertex of
degree $1$.  Define the partition function of these pulled stars as:
\begin{equation}
S_n^{(f)}(a,y) = \sum_{v,h} s_n^{(f)}(v,h) \, a^v y^h.
\label{eqn:starPF}
\end{equation}
If the star is instead pulled at its central node of degree $f$ then we need to
keep track of the height of this central vertex.  Write $\widehat{s}_n^{(f)}(v,h)$ 
for the number of $f$-stars with $v+1$ vertices in $x_3=0$ and with the central
node at height $h$.  Define the partition function
\begin{equation}
\widehat{S}_n^{(f)}(a,y) = \sum_{v,h} \widehat{s}_n^{(f)}(v,h) \, a^v y^h.
\label{eqn:starcentralPF}
\end{equation}

\subsection{Adsorbing pulled unfolded bridges in wedges}
\label{sec:bridges}

Throughout the remainder of the paper we shall find it convenient to deal with 
unfolded walks \cite{HammersleyWelsh} and with walks confined to wedges
\cite{HammersleyWhittington}.  These subsets of self-avoiding walks have the useful property 
that they have the same cardinality, to exponential order,  as the set of all self-avoiding walks.  
A \textit{bridge} is a particular kind of unfolded walk \cite{HammersleyWelsh}, and in
this section we will introduce a modified bridge called an $\alpha$-bridge which is 
unfolded in the $x_1$ and $x_2$ directions and contained in a wedge in the lattice.

\begin{figure}[t]
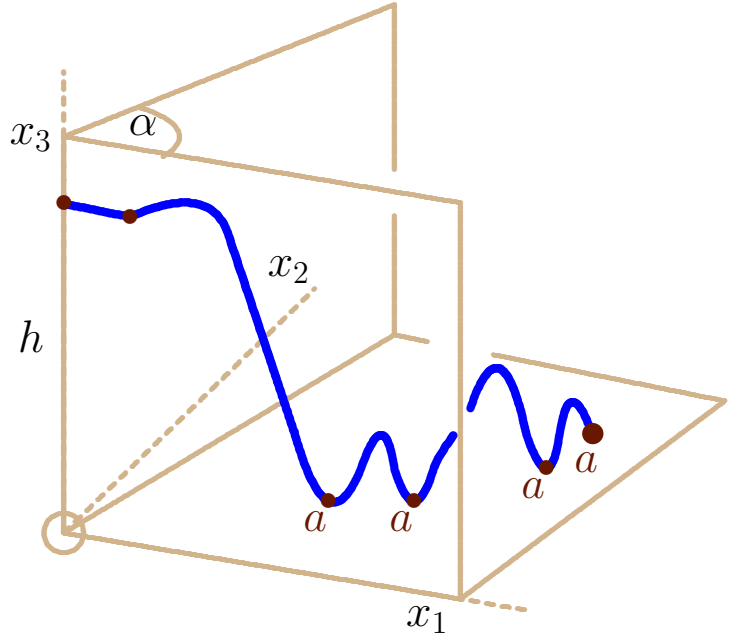

\beginpicture
\setcoordinatesystem units <2.5pt,2.5pt>
\setplotarea x from -40 to 160, y from -15 to 80
\setplotarea x from 0 to 150, y from 0 to 70

\color{Tan}
\setplotsymbol ({\LARGE$\cdot$})

\setdashes <3pt>
\plot 0 0 70 -11.67 /
\plot 0 0 38 37 /
\plot 0 0 0 70 /

\setsolid
\plot 60 -10 0 0 50 30 50 48 /
\plot 50 55 50 80 0 60 60 50 60 -10 /
\plot 0 0 0 60 /
\plot 60 -10 100 20 65 27 /
\plot 55 29 50 30 /

\circulararc 360 degrees from 0 3 center at 0 0 

\ellipticalarc axes ratio 3:1 80 degrees from 15 57 center at 0 60 

\color{black}
\put {\LARGE$\alpha$} at 12 62 

\put {\LARGE$x_1$} at 55 -13
\put {\LARGE$x_2$} at 34 40
\put {\LARGE$x_3$} at -5 60
\put {\LARGE$h$} at -5 30

\setplotsymbol ({\scriptsize$\bullet$})
\color{Blue}
\plot 0 50 10 48 /

\setquadratic
\plot 10 48 20 50 25 45 30 30 35 15 40 5 45 10 48 15 50 10 53 5 56 10 57 12 59 15 /
\plot 61.5 19 66 25 70 15 73 10 75 15 77 20 80 15 /

\color{Sepia}
\multiput {\Large$\bullet$} at 0 50 10 48 40 5 53 5 73 10  /
\multiput {\LARGE $a$} at 38 2 51 2 71 7  79 11 /
\color{Sepia}
\multiput {\huge$\bullet$} at 80 15 /

\color{black}
\normalcolor
\endpicture
\caption{An adsorbing pulled arm of an $f$-star in an $\alpha$-wedge.  Vertices in 
the arm adsorb in the adsorbing plane with activity $a$. }
\label{figure2} 
 \end{figure}

In figure \ref{figure2} we show a schematic diagram of an arm of an $f$-star in a 
half-wedge of angle $\alpha$ and with \textit{floor} the $(x_1,x_2)$-plane.   
The half-wedge is bounded by its floor and by two planes (the first the $(x_1,x_3)$-plane, 
and the second a plane through the $x_3$-axis at an angle $\alpha$ with the first).  The 
two bounding planes of the wedge meet in the positive $x_3$-axis, which is the
\textit{spine} of the wedge.  Generally, we assume
that the angle $\alpha< \sfrac{\pi}{2}$, and later on we shall put $\alpha = \sfrac{\pi}{4}$.
These half-wedges will be called \textit{$\alpha$-wedges}.

A \textit{$\alpha$-bridge} is a doubly unfolded walk in the $\alpha$-wedge 
with an end-vertex in the spine of the wedge at height $h$ above 
the floor.  The $\alpha$-bridge
\begin{enumerate}
\item is unfolded \cite{HammersleyWelsh} in the $x_1$- and the $x_2$-directions, and 
\item its terminal vertex is in the floor of the wedge.
\end{enumerate}
In addition to being confined to the $\alpha$-wedge this means that it satisfies the conditions:
\begin{enumerate}
\item
$0= x_1(0) < x_1(k) \le x_1(n)$ for $0<k<n$,
\item
$0=x_2(0) \le x_2(k) \le x_2(n)$ for $0<k<n$, and
\item
$x_3(0)=h$ and $x_3(n)=0$.
\end{enumerate} 
Let the number of these $\alpha$-bridges of length $n$
with $v$ visits to the floor to the $\alpha$-wedge, and with height of first vertex $h$, be $b_n^{(\alpha)}(v,h)$.  The
partition function of these $\alpha$-bridges is
\begin{equation}
B_n^{(\alpha)}(a,y) = \sum_{v,h} b_n^{(\alpha)}(v,h)\, a^v y^h .
\end{equation}
Our aim is to concatenate bridges in a wedge, as illustrated schematically in Figure 2 (b), in
order to build up an $\alpha$-bridge in an $\alpha$-wedge.  We proceed 
by using the approach in reference \cite{HammersleyWhittington}
(see also section 8.5 in reference \cite{Rensburg2015} for a similar approach
applied to lattice polygons).

Let $b_m(h)$ be the number of \textit{doubly unfolded bridges} (in the
$x_1$ and $x_2$ directions) of length $m$ and height $h$.
Then the partition function of these doubly unfolded bridges is
\begin{equation}
B_m(y) = \sum_{h=0}^m b_m(h)\, y^h .
\end{equation}
The \textit{displacement vector} $\vec{\delta}$ of bridges contributing to $B_m(y)$ is the
difference between the endpoints of the bridge.  In three dimensions there are at 
most $8(m+1)^3$ different displacement vectors, and there is a most popular
displacement vector $\vec{\delta}^*$.  That is, if there are $b_m^*(h)$ bridges with
the most popular displacement vector $\vec{\delta}^*$, then 
\begin{equation}
B_m^* (y) =
\sum_m b_m^* (h)\, y^h \leq B_m(y) \leq 8(m+1)^3 \sum_m b_m^*(h)\, y^h .
\end{equation}
The free energy of these bridges pulled in the $x_3$ direction is $\lambda(y)$,
since unfolding a walk in the $x_1$ and $x_2$ directions does not change the
$x_3$ coordinate of any vertex.  This shows that
\begin{equation}
\lambda(y) = \lim_{m\to\infty} \sfrac{1}{m} \log B_m(y) 
= \lim_{m\to\infty} \sfrac{1}{m} \log \sum_m b_m^* (h)\, y^h 
= \lim_{m\to\infty} \sfrac{1}{m} \log B_m^*(y) .
\label{eqn13A}
\end{equation}
We assume, without loss of generality and by symmetry, that $\vec{\delta}^*$ is a vector 
in the first octant in the cubic lattice, and that its reflection in the $(x_1,x_3)$-plane is 
$\vec{\delta}^\bullet$ which is a vector in the fourth octant.  Denote the class of bridges
with displacement vector $\vec{\delta}^\bullet$ by $b_m^\bullet(h)$ and denote the
partition function of these bridges by $B_m^\bullet(y)$.  

Observe that $b_m^\bullet(h) = b_m^*(h)$.  Then $B_m^*(y) = B_m^\bullet(y)$, and
bridges from $B_m^*(y)$ and $B_m^\bullet(y)$ with
displacement vectors $\vec{\delta}^*$ and $\vec{\delta}^\bullet$ can be
concatenated to obtain bridges unfolded in the $x_1$ direction.

\begin{figure}[t]
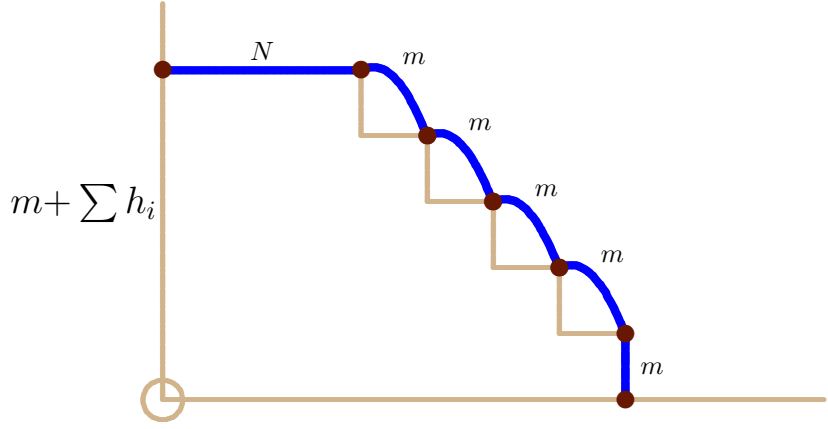

\beginpicture
\setcoordinatesystem units <2.5pt,2.5pt>
\setplotarea x from -40 to 160, y from -15 to 80
\setplotarea x from 0 to 150, y from 0 to 70

\color{Tan}
\setplotsymbol ({\LARGE$\cdot$})

\setsolid
\plot 0 60 0 0 100 0 /

\circulararc 360 degrees from 0 3 center at 0 0

\color{black}
\put {\Large$m{+}\sum h_i$} at -12 30

\setplotsymbol ({\scriptsize$\bullet$})
\color{Blue}
\plot 0 50 30 50 /
\plot 70 0 70 10 /

\multiput 
{\beginpicture
\setquadratic
\setplotsymbol ({\scriptsize$\bullet$})
\color{Blue}
\plot  10 10 15 9 20 0 /
\setlinear
\setplotsymbol ({\LARGE$\cdot$})
\color{Tan}
\plot 10 10 10 0 20 0 /
\color{Sepia}
\multiput {\LARGE$\bullet$} at 10 10 20 0 /
\endpicture}
at 35 45 45 35 55 25 65 15 
/
\put {\LARGE$\bullet$} at 0 50 
\put {\LARGE$\bullet$} at 70 0

\color{black}
\put {$N$} at 15 53
\multiput {$m$} at 38 52 48 42 58 32 68 22 74 5 
/

\color{black}
\normalcolor
\endpicture
\caption{Schematic drawing of $k-1$ adsorbing pulled doubly unfolded bridges of length $m$
each stacked in an $\alpha$-wedge. These bridges are joined to the spine of the wedge by
a line segment of $N$ edges, and to the floor of the wedge by a line segment of length $m$.}
\label{figure3} 
 \end{figure}

Using the methods of reference \cite{HammersleyWhittington}
an arbitrary number $k$ of bridges of length $m$ with most popular
displacement vectors $\vec{\delta}^*$ and $\vec{\delta}^\bullet$ can be concatenated
in an $\alpha$-wedge and joined to the spine of the wedge by a string of $N$ edges,
as illustrated in figure \ref{figure3}.  See, for example, section 8.5 in reference
\cite{Rensburg2015} for polygons in a wedge. Since $B_m^*(y)$ and 
$B_m^\bullet(y)$ are the most popular classes of these doubly unfolded bridges,
let $b_m^*(h_i)$ and $b_m^\bullet(h_i)$ be the number of bridges of height 
$h_i$ contributing to $B_m^*(y)$ and $B_m^\bullet(y)$, respectively.  Since
$b_m^\bullet(h) = b_m^*(h)$, this shows that the partition function of 
$\alpha$-bridges is bounded from below:
\begin{equation}
\hspace{-1cm}
B_{N+km}^{(\alpha)} (1,y) 
= \sum_h b_{N+km}^{(\alpha)} (h)\, y^h 
\geq \sum_{\{h_i\}}\prod_{i=1}^{k-1} b_m^*(h_i)\, y^{m+\sum_i h_i}
= \L B_m^*(y) \R^{k-1}\,y^m .
\end{equation}
Let $n=N+km$, take the logarithm and divide by $n$.  Take the liminf as 
$n\to\infty$ on the left hand side with
$m$ fixed and with $L\leq N \leq L+m$ for a large fixed $L\gg m$.  Then $k\to\infty$
and it follows that
\begin{equation}
\liminf_{n\to\infty} \sfrac{1}{n} \log B_n^{(\alpha)}(y) \geq \sfrac{1}{m} \log B_m^* (y).
\label{eqn12}
\end{equation}
By equation \Ref{eqn13A}
this gives the following lemma.

\begin{lemm}
$ \displaystyle \lim_{n\to\infty} \sfrac{1}{n} \log  B_n^{(\alpha)}(1,y) = \lambda (y) .$
\end{lemm}

\Pr  Since $B_n^{(\alpha)}(1,y) \leq B_n(y)$ the result follows since, by equation \Ref{eqn12},  
\[ \hspace{-2cm}
\liminf_{n\to\infty} \Sfrac{1}{n} \log B_n^{(\alpha)}(1,y) \geq
\lim_{m\to\infty}  \Sfrac{1}{m} \log B_m^* (y) =
\lim_{m\to\infty}  \Sfrac{1}{m} \log B_m (y) \geq
\limsup_{m\to\infty} \Sfrac{1}{m} \log B_m^{(\alpha)}(1,y) 
\]
By equation \Ref{eqn13A} this
completes the proof. \qed

We shall also need to work with $\alpha$-loops in an $\alpha$-wedge.  These are defined
as $\alpha$-bridges with first vertex at the origin in the $\alpha$-wedge and terminal
(last) vertex in the floor of the wedge.  If the number 
of such loops of length $n$ with $v$ visits (excluding the vertex at the origin) is given by
$\ell_n^{(\alpha)}(v)$, then $\ell_n^{(\alpha)}(v) = b_n^{(\alpha)}(v,0)$
and the partition function of $\alpha$-loops is given by
\begin{equation}
L_n^{(\alpha)} (a) = \sum_v \ell_n^{(\alpha)}(v) \, a^v .
\end{equation}

 We now similarly consider loops in an $\alpha$-wedge.  Let $L_n^\ddagger(a)$ be the number
\textit{doubly unfolded loops} (in the $x_1x_2$-directions) from the origin in the cubic lattice.  
Any loop of length $n$ can be fitted into an $\alpha$-wedge, provided that it is 
placed far enough from the origin.  Suppose that $N$ is a large and fixed number, such that any
loop of length $n$ can be placed in the $\alpha$-wedge, and joined to the origin with
$N$ edges.

\begin{figure}[t]
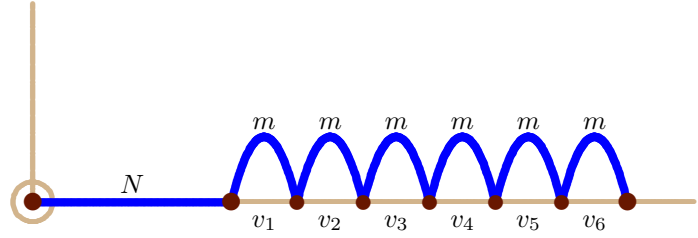

\beginpicture
\setcoordinatesystem units <2.5pt,2.5pt>
\setplotarea x from -40 to 160, y from -10 to 35
\setplotarea x from 0 to 150, y from 0 to 30

\color{Tan}
\setplotsymbol ({\LARGE$\cdot$})

\setsolid
\plot 0 30 0 0 100 0 /

\circulararc 360 degrees from 0 3 center at 0 0 

\color{black}

\setplotsymbol ({\scriptsize$\bullet$})
\color{Blue}
\plot 0 0 30 0 /

\multiput 
{\beginpicture
\setquadratic
\setplotsymbol ({\scriptsize$\bullet$})
\color{Blue}
\plot  0 0 5 10 10 0  /
\setlinear
\setplotsymbol ({\LARGE$\cdot$})
\color{Sepia}
\multiput {\Large$\bullet$} at 0 0  /
\endpicture}
at  30 0 40 0 50 0 60 0 70 0 80 0   /
\color{Sepia}
\put {\LARGE$\bullet$} at 0 0 

\color{Sepia}
\multiput {\LARGE$\bullet$} at 30 0  90 0   /

\color{black}
\put {$N$} at 15 3
\multiput {$m$} at 35 12 45 12 55 12 65 12 75 12 85 12 /
\put {$v_1$} at 35 -3  \put {$v_2$} at 45 -3  \put {$v_3$} at 55 -3 
 \put {$v_4$} at 65 -3  \put {$v_5$} at 75 -3  \put {$v_6$} at 85 -3

\color{black}
\normalcolor
\endpicture
\caption{Schematic diagram of adsorbing doubly unfolded loops concatenated in
an $\alpha$-wedge. }
\label{figure4} 
 \end{figure}

Then it is possible to use most popular class arguments to concatenate $m$ loops
in a sequence inside the $\alpha$-wedge to create an  $\alpha$-loop of
length $N+km$.  This is illustrated schematically in figure \ref{figure4}.   The details of
the construction are similar to that explained for walks or polygons in wedges
(see, for example, section 8.5 in reference \cite{Rensburg2015} for a polygons in a wedge,
and reference \cite{HammersleyWhittington} for walks in a wedge).  While the particular details
are slightly different, the construction is \emph{mutatis mutandis}, and the result is
the following inequality
\begin{equation}
L_{N+km}^{(\alpha)} (a) \geq a^N (L_m^{\ddagger,*}(a))^k ,
\end{equation}
where $L_m^{\ddagger,*}(a)$ is the partition function of a most popular class of
doubly unfolded loops with the property that $\lim_{n\to\infty} \sfrac{1}{n}
\log L_n^{\ddagger,*}(a) = \kappa(a)$.  $L_n^{(\alpha)}(a)$ is the partition function 
of loops from the spine of the $\alpha$-wedge and
contained inside the $\alpha$-wedge.  Take logarithms, divide by $N+km$ and take 
the limit inferior on the left hand side as $k\to\infty$ with $m$ fixed.  If the liminf 
is realised along a subsequence $N_i+k_i m$, where $N+m> N_i\geq N$, then $k_i \to\infty$, 
and this shows that 
\begin{equation}
\liminf_{n\to\infty} \Sfrac{1}{n} \log L_n^{(\alpha)} (a) \geq \Sfrac{1}{m} \log L_m^\ddagger (a).
\end{equation}
Taking the limit on the right hand side as $m\to\infty$
gives $\liminf_{n\to\infty}\Sfrac{1}{n} \log L_n^{(\alpha)} (a) \geq
\kappa(a)$, and so this gives the following lemma for adsorbing $\alpha$-loops:

\begin{lemm}
$ \displaystyle \lim_{n\to\infty} \sfrac{1}{n} \log L_n^{(\alpha)}(a) = \kappa (a) .$
\end{lemm}

\Pr  By inclusion $\limsup_{n\to\infty} \sfrac{1}{n} \log L_n^{(\alpha)}(a) \leq \kappa (a)$,
so the theorem follows. \qed

An adsorbing $\alpha$-bridge in an $\alpha$-wedge has its first visit in the floor of the 
wedge (see figure \ref{figure2}) whereafter it continues as a loop in the wedge.  If we
assume this is an $\alpha$-loop in an $\alpha$-wedge, then a lower bound is obtained:
\begin{equation}
B_n^{(\alpha)}(a,y) \geq a \sum_{hv} b_{m^*}^{(\alpha)}\, \ell_n^{(\alpha)}(v)\, a^v y^h
= a\, B_{m^*}^{(\alpha)}(y)\, L_{n-m^*-1}^{(\alpha)} (a) ,
\end{equation} 
where $m^* \in\{0,n\}$ (and $m^*$ is a function of $n$).  Choosing $m^* = \lfloor \zeta n \rfloor$,
taking logs, dividing by $n$ and taking $n\to\infty$ gives
\begin{equation}
\liminf_{n\to\infty} \sfrac{1}{n} \log B_n^{(\alpha)} (a,y) \geq
\zeta \lambda(y) + (1-\zeta) \kappa (a) .
\end{equation}
Taking the maximum over $\zeta$ gives
\begin{equation}
\liminf_{n\to\infty} \sfrac{1}{n} \log B_n^{(\alpha)} (a,y) \geq
\max\{\kappa(a),\lambda(y)\} = \psi(a,y).
\label{eqn17}  
\end{equation}
This result gives the following theorem.

\begin{theo}
For adsorbing pulled $\alpha$-bridges 
$ \displaystyle \lim_{n\to\infty} \sfrac{1}{n} \log B_n^{(\alpha)}(a,y) = \psi(a,y) .$
\label{theorem1} 
\end{theo}

\Pr
Observe that every $\alpha$-bridge in an $\alpha$-wedge has a first visit to the floor where
it can be cut to give a pulled walk of length $m$ and an adsorbing loop.  That is,
\[ B_n^{(\alpha)}(a,y) \leq \sum_m C_m^+ (y)\, L_{n-m-1}(a). \]
where $C_m^+(y) = C_m^+(1,y)$.  This shows that
\[ \limsup_{n\to\infty} \sfrac{1}{n} \log B_n^{(\alpha)} (a,y) \leq
\max\{\kappa(a),\lambda(y)\} = \psi(a,y). \]
Together with equation \Ref{eqn17} this completes the proof. \qed

\subsection{Pulled Spiders}

\begin{figure}[t]
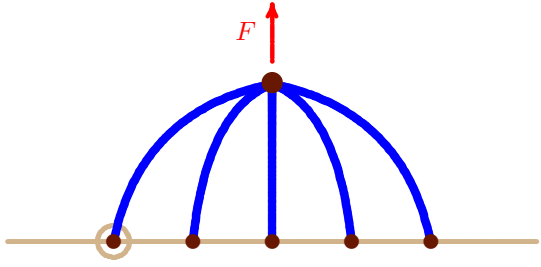

\beginpicture
\setcoordinatesystem units <2.0pt,2.0pt>
\setplotarea x from -50 to 160, y from -5 to 45
\setplotarea x from 0 to 150, y from 0 to 30

\color{Tan}
\setplotsymbol ({\LARGE$\cdot$})

\setsolid
\plot 0 0 100 0 /

\circulararc 360 degrees from 20 3 center at 20 0 

\color{black}

\setplotsymbol ({\scriptsize$\bullet$})
\color{Blue}
\setquadratic
\plot 50 30 30 20 20 0  /
\plot 50 30 40 20 35 0 /
\plot 50 30 50 20 50 0 /
\plot 50 30 60 20 65 0 /
\plot 50 30 70 20 80 0 /

\color{Sepia}
\multiput {\Large$\bullet$} at 20 0 35 0 50 0 65 0 80 0    /

\put {\huge$\bullet$} at 50 30 

\setplotsymbol ({\Large$\cdot$})
\color{red}
\arrow  <5pt> [0.2,.67] from 50 34 to  50 45
\put {$F$} at 45 40

\color{black}
\normalcolor
\setlinear

%
%
%
%
%

%
%
%
%
%

\endpicture
\caption{A schematic diagram of a pulled spider.
}
\label{figure5} 
 \end{figure}

A \textit{spider} is a uniform $f$-star with one end-vertex at the origin, and each
arm has at least one visit to the adsorbing plane, and it is pulled with a force $F$
at its central node.  A spider is schematically illustrated in figure \ref{figure5}.

\begin{figure}[t]
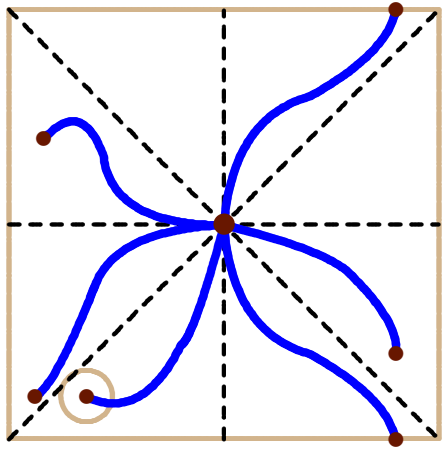

\beginpicture
\setcoordinatesystem units <3.25pt,3.25pt>
\setplotarea x from -35 to 160, y from -5 to 55
\setplotarea x from 0 to 150, y from 0 to 50

\color{Tan}
\setplotsymbol ({\LARGE$\cdot$})

\setsolid
\plot 0 0 50 0 50 50 0 50 0 0  /

\circulararc 360 degrees from 9 8 center at 9 5  

\setplotsymbol ({\Large$\cdot$})
\color{black}
\setdashes <4pt>
\plot 0 0 50 50 /
\plot 0 50 50 0 /
\plot 0 25 50 25 /
\plot 25 0 25 50 /

\setsolid
\setplotsymbol ({\scriptsize$\bullet$})
\color{Blue}
\setquadratic
\plot 25 25 28 35 35 40 42 45 45 50 /
\plot 25 25 28 15 35 10 42 5 45 0 /
\plot 25 25 22 15 20 11 15 5 9 5  /
\plot 25 25 15 23 10 18 6 9 3 5 /
\plot 25 25 35 22 40 19 44 14 45 10 /
\plot 25 25 15 27 11 33 8 37 4 35 /

\color{Sepia}
\multiput {\Large$\bullet$} at 45 50 45 0 9 5 3 5 45 10 4 35    /

\put {\huge$\bullet$} at 25 25

\color{black}
\normalcolor
\endpicture
\caption{Top view of an adsorbing pulled spider made from adsorbing bridges in $\alpha$-wedges. }
\label{figure6} 
 \end{figure}

A pulled spider with adsorbing legs can be created from unfolded bridges in $\alpha$-wedges
with $\alpha = \sfrac{\pi}{4}$.  A top view of a $6$-star is shown in figure \ref{figure6}, and the
arrangement around the central node is as illustrated in figure \ref{figure1}.  

The partition function of pulled $f$-spiders is denoted by
$U_n^{(f)} (a,y)$.  We now outline a proof that the limiting free energy of
an $f$-spider is $\psi(a,y^{1/f})$.  

Putting together bridges in $\alpha$-wedges as illustrated in figure \ref{figure6}
gives the lower bound
\begin{equation}
U_{fn}^{(f)} (a,y) \geq
\sum_h \L \sum_v b_{n-k}^{(\pi/4)}(v,h)\, a^v \R^{\! f} y^h 
\end{equation}
where $k$ is small and fixed and accounts for putting together the bridges at the central node
as shown in figure \ref{figure1}.  In the summations on the right there is a most popular value
of $h$, say $h^*$ (a function of $(n,k,a,y)$).  This shows shows that
\begin{eqnarray}
U_{fn}^{(f)} (a,y) 
&\geq  \L \sum_v b_{n-k}^{(\pi/4)}(v,h^*)\, a^v \R^{\! f} y^{h^*}  \nonumber \\
&\geq \L \Sfrac{1}{n+1} \sum_{v,h} b_{n-k}^{(\pi/4)}(v,h)\, a^v y^{h/f} \R^{\! f} 
= \L \Sfrac{1}{n+1} \, B_{n-k}^{(\pi/4)} (a,y^{1/f}) \R^{\! f} .
\end{eqnarray}
Take logarithms, divide by $nf$, and let $n\to\infty$.  This shows that
\begin{equation}
\liminf_{n\to\infty} \Sfrac{1}{fn} \log U_{fn}^{(f)} (a,y)
\geq \max\{\kappa(a), \lambda(y^{1/f}) \} = \psi(a,y^{1/f}).
\label{eqn19}  
\end{equation}
by theorem \ref{theorem1}.

On the other hand, every spider can be decomposed by cutting it into $f$ walks
in the central node.   Each such walk starts at height $h$ in the central node,
makes a first visit to the adsorbing plane, whereafter it continues as a positive adsorbing
walk.

 This shows that
\begin{eqnarray}
U_{fn}^{(f)}(a,y) 
&\leq \sum_h \L \sum_k c_{k}^+ (1,h)
\L \sum_v c_{n-k}^+(v,\bullet)\, a^v \R  \R^{\! f} y^h \\
&\leq  \L \sum_k \L \sum_h c_k^+(1,h)\, y^{h/f} \R
\L \sum_{v} c_{n-k}^+(v,\bullet)\, a^v \R \R^{\! f} .
\end{eqnarray}
where $c_n^+(v,\bullet) = \sum_h c_n^+(v,h)$.  Taking the logarithm, dividing by
$fn$ and taking $n\to\infty$ gives
\begin{equation}
\limsup_{n\to\infty} \Sfrac{1}{fn} \log U_{fn}^{(f)}(a,y) 
\leq \max\{ \kappa(a), \lambda( y^{1/f}) \} = \psi(a,y^{1/f}) .
\end{equation}
This gives the following theorem.

\begin{theo} 
For pulled adsorbing spiders 
$ \displaystyle \lim_{n\to\infty} \sfrac{1}{fn} \log U_{fn}^{(f)}(a,y) = \psi(a,y^{1/f}) .$ \qed
\label{theorem2}  
\end{theo}


\subsection{Uniform $f$-stars pulled at the central vertex}
\label{sec:starscentral}

In this section we examine uniform adsorbing $f$-stars  pulled at their central node (see
figure \ref{figure1}).  Each arm of the star may have visits in the adsorbing plane, but only
one branch has an end-vertex fixed in the origin.  We call these $\widehat{S}$-stars.
The partition function of an $\widehat{S}$-star with $f$ arms is
$\widehat{S}^{(f)}_n (a,y)$ (see equation \Ref{eqn:starcentralPF}). 

\begin{figure}[t]
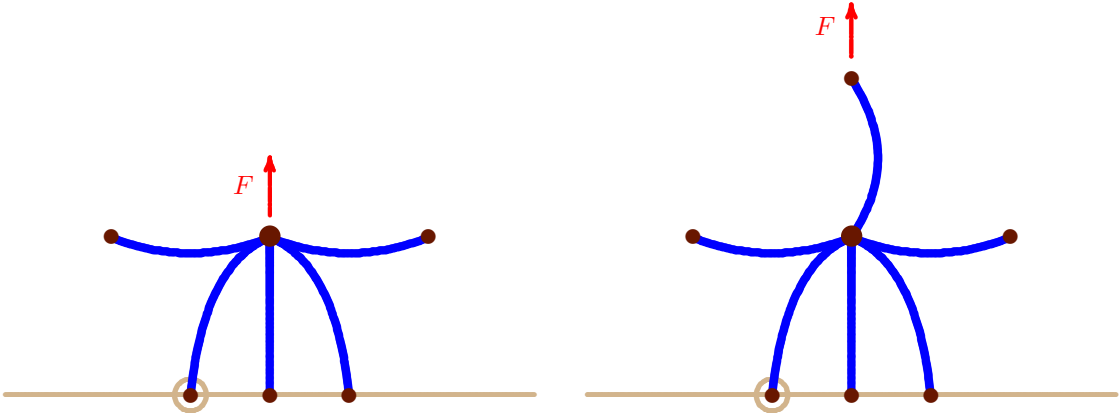

\beginpicture
\setcoordinatesystem units <2.0pt,2.0pt>
\setplotarea x from 0 to 160, y from -5 to 45
\setplotarea x from 0 to 150, y from 0 to 30

\color{Tan}
\setplotsymbol ({\LARGE$\cdot$})

\setsolid
\plot 0 0 100 0 /

\circulararc 360 degrees from 35 3 center at 35 0 

\color{black}

\setplotsymbol ({\scriptsize$\bullet$})
\color{Blue}
\setquadratic
\plot 50 30 35 27 20 30  /
\plot 50 30 40 20 35 0 /
\plot 50 30 50 20 50 0 /
\plot 50 30 60 20 65 0 /
\plot 50 30 65 27 80 30 /

\color{Sepia}
\multiput {\Large$\bullet$} at 20 30 35 0 50 0 65 0 80 30    /

\put {\huge$\bullet$} at 50 30 

\setplotsymbol ({\Large$\cdot$})
\color{red}
\arrow  <5pt> [0.2,.67] from 50 34 to  50 45
\put {$F$} at 45 40

\color{black}
\normalcolor
\setlinear

\setcoordinatesystem units <2.0pt,2.0pt> point at -110 0 
\setplotarea x from  0 to 160, y from -5 to 45
\setplotarea x from 0 to 150, y from 0 to 75

\color{Tan}
\setplotsymbol ({\LARGE$\cdot$})

\setsolid
\plot 0 0 100 0 /

\circulararc 360 degrees from 35 3 center at 35 0 

\color{black}

\setplotsymbol ({\scriptsize$\bullet$})
\color{Blue}
\setquadratic
\plot 50 30 35 27 20 30  /
\plot 50 30 40 20 35 0 /
\plot 50 30 50 20 50 0 /
\plot 50 30 60 20 65 0 /
\plot 50 30 65 27 80 30 /

\setquadratic
\plot 50 30 55 45 50 60  /

\color{Sepia}
\multiput {\Large$\bullet$} at 20 30 35 0 50 0 65 0 80 30 50 60  /

\put {\huge$\bullet$} at 50 30 

\setplotsymbol ({\Large$\cdot$})
\color{red}
\arrow  <5pt> [0.2,.67] from 50 64 to  50 74
\put {$F$} at 45 70

\color{black}
\normalcolor

\endpicture
\caption{A schematic diagram of a pulled $\widehat{S}$-star (left), and a pulled $S$-star (right). }
\label{figure7} 
 \end{figure}

Each $\widehat{S}$-star has the structure shown on the left in figure \ref{figure7}.
The star has $f$ arms, and it is pulled by a force $F$ in its central node.  Including
the arm with its end-vertex at the origin, there are $g$ arms which have visits to 
the adsorbing plane, and $f-g$ arms which are disjoint with the adsorbing plane.
In other words, the structure of the $\widehat{S}$-star is that of a pulled
spider with $g$ arms, and with $f-g$ arms appended in the central node and disjoint
with the adsorbing plane. 

A lower bound is obtained on the free energy by using the arguments leading to 
equation \Ref{eqn19}.  Each of the $g$ arms of the spider are taken to be
$\alpha$-bridges (in disjoint $\alpha$-wedges as shown in figure
\ref{figure6}).  The remaining $f-g$ arms are self-avoiding walks confined to $\alpha$-wedges
and are disjoint with the adsorbing floors of the infinite wedges.  This shows that
\begin{equation}
\liminf_{n\to\infty} \Sfrac{1}{fn} \log \widehat{S}^{(f)}_{fn} (a,y)
\geq \Sfrac{g}{f}\, \psi(a,y^{1/g}) + \Sfrac{f-g}{f}\, \log \mu_3 .
\end{equation}
Notice that $g\geq 1$ since at least one  arm has a visit in the adsorbing plane.
If $\kappa(a)>\lambda(y)$ then the right hand side is maximized when $g=f$ and
 $\psi(a,y^{1/f}) = \kappa(a)$ (since $\lambda(y^{1/f}) \leq \lambda(y)$).
This shows that $\liminf_{n\to\infty} \Sfrac{1}{fn} \log \widehat{S}^{(f)}_{fn} (a,y)
\geq \kappa(a)$.

On the other hand, if $\kappa(a) < \lambda(y)$, then the maximum is achieved
with $g=1$.  To see this, notice that since $\lambda(y)$ non-decreasing with $y$,
$\lambda(y) \geq \lambda(y^{1/g})$ if $y \geq 1$, and 
\begin{equation}
\lambda(y) + (g-1) \log \mu_3
\geq g \, \lambda(y^{1/g})
\end{equation}
since $\lambda(y)$ is a convex function of $\log y$.  Thus
\begin{equation}
 \Sfrac{g}{f}\, \lambda(y^{1/g}) + \Sfrac{f-g}{f}\, \log \mu_3
\leq \Sfrac{1}{f} \, \lambda(y) + \Sfrac{f-1}{f}\, \log \mu_3 
\label{eqn26}  
\end{equation}
for any $g \in \{ 1,2,\ldots,f-1\}$. This shows that $\liminf_{n\to\infty} \Sfrac{1}{fn} 
\log \widehat{S}^{(f)}_{fn} (a,y) \geq \Sfrac{1}{f} \,\lambda(y) + \Sfrac{f-1}{f}\,
\log \mu_3$.  The result is that
\begin{equation}
\liminf_{n\to\infty} \Sfrac{1}{fn} \log \widehat{S}^{(f)}_{fn} (a,y)
\geq \max\LC  \kappa(a), \Sfrac{1}{f} \,\lambda(y) + \Sfrac{f-1}{f}\, \log \mu_3 \RC.
\end{equation}

An upper bound is obtained by considering the spider in the star, and the remaining
arms to be independent.  If the spider has $g$ legs, then its contribution is
$\sfrac{g}{f}\, \psi(a,y^{1/g})$.  Each remaining arm is disjoint with the adsorbing
plane (if it is not, then it will be part of the spider), and so contributes
$\sfrac{1}{f}\,\log \mu_3$.  This gives the upper bound
\begin{equation}
\limsup_{n\to\infty} \Sfrac{1}{fn} \log \widehat{S}^{(f)}_{fn} (a,y)
\leq \max_{g\geq 1}
\L  \Sfrac{g}{f}\, \psi(a,y^{1/g})  + \Sfrac{f-g}{f} \, \log \mu_3 \R .
\end{equation}
As above, if $\kappa(a) > \lambda(y)$ then the maximum is obtained when
$g=f$ and is equal to $\kappa(a)$, and if $\kappa(a) < \lambda(y)$ when
$g=1$.  Put this together to complete the proof of the following theorem.

\begin{theo}
$\displaystyle
\widehat{\sigma}^{(f)} (a,y)= \lim_{n\to\infty} \Sfrac{1}{fn} \log  \widehat{S}^{(f)}_{fn} (a,y)
= \max\LC \kappa(a), \sfrac{1}{f}\, \lambda(y) + \sfrac{f-1}{f}\, \log \mu_3 \RC$. \qed
\end{theo}

This shows that there is a phase boundary in this model at the solution of
\begin{equation}
\lambda(y) = f\, \kappa(a) - (f-1)\log \mu_3.
\end{equation}
This phase boundary is first order, and it separates an adsorbed phase with
$\widehat{\sigma}^{(f)}(a,y) =\kappa(a)$ from a ballistic phase with
$\widehat{\sigma}^{(f)} (a,y)= \sfrac{1}{f}\, \lambda(y) + \sfrac{f-1}{f}\, \log \mu_3$
(that is, one arm is ballistic, and the remaining $f-1$ arms are free). 

For example, if $f=1$ then this is a pulled adsorbing walk, and
$\widehat{\sigma}^{(1)} (a,y)= \max\{ \kappa(a),\lambda(y)\} = \psi(a,y)$, as
expected, and the phase boundary is given by the solution of
$\kappa(a) = \lambda(y)$; see theorem 1 in reference \cite{Rensburg2016b}.

If $f=2$, then this is a model of a walk pulled in its middle vertex.  This 
gives $\widehat{\sigma}^{(2)} (a,y)= \max\{ \kappa(a),\sfrac{1}{2}(\lambda(y)+\log \mu_3)\}$
(see reference \cite{Rensburg2017}).  The phase boundary is given by the solution of
$2\,\kappa(a) = \lambda(y) + \log \mu_3$ (see figure 6 in reference \cite{Rensburg2017}).

If $f=3$, then the result is that 
$\widehat{\sigma}^{(3)}(a,y) = \max\{ \kappa(a),\sfrac{1}{3}(\lambda(y)+2\log \mu_3)\}$,
and the phase boundary is given by $3\,\kappa(a) = \lambda(y)+2\log \mu_3$ -- 
this shows an adsorbed phase, and a ballistic phase where the arm attached to
the adsorbing surface is ballistic, and the other two arms are free.  In other words,
the unattached arms both desorb at the same time.  This generalises to general
values of $f>2$ with the phase boundary given by
$f\thin \kappa(a) = \lambda(y)+(f-1)\log \mu_3$.

\subsection{Uniform $f$-stars pulled at the end-vertex of an arm}
\label{sec:starsendvertex}

In this section we examine uniform $f$-stars adsorbed at a surface and pulled at a
vertex of degree 1.  We call this model $S$-stars.  This model was already  considered 
for 3-stars \cite{Rensburg2018} and we generalize it here to the case
$3 \le f \le 6$.  The interesting question is whether more phases occur for larger values of $f$.

An $S$-star is illustrated on the right in figure \ref{figure7}.  The star is composed of
a spider with $g$ legs, one arm pulled at its end-vertex, and $f-g-1$ arms which are 
disjoint with the adsorbing plane.  The partition function of the star is given 
by $S_{fn}^{(f)}(a,y)$ (see equation \Ref{eqn:starPF}).

A lower bound is obtained by restricting each arm of the star to an $\alpha$-wedge
(with top view as illustrated in figure \ref{figure6}).  The contribution of the spider 
to the lower bound is $\sfrac{g}{f} \, \psi(a,y^{1/g})$ by equation \Ref{eqn19}.
The pulled arm contributes $\sfrac{1}{f}\, \psi(a,y)$ and the remaining $f-g-1$
arms $\sfrac{f-g-1}{f}\, \log \mu_3$.  Putting this together gives
\begin{equation}
\liminf_{n\to\infty} \Sfrac{1}{fn} \log S_{fn}^{(f)} (a,y) \geq
\Sfrac{g}{f} \, \psi(a,y^{1/g}) + \Sfrac{1}{f} \, \psi(a,y) + \Sfrac{f-g-1}{f} \, \log \mu_3 .
\label{eqn29}   
\end{equation}

If $a \leq a_c$ then $\psi(a,y^{1/g}) = \lambda(y^{1/g})$ and $\psi(a,y) = \lambda(y)$.
The right hand side is then a maximum if $g=1$ by equation \Ref{eqn26} in
which case $\liminf_{n\to\infty} \Sfrac{1}{fn} \log S_{fn}^{(f)} (a,y) \geq
\sfrac{2}{f}\, \lambda(y) + \sfrac{f-2}{f} \, \log\mu_3$.

If, on the other hand, $y<1$, then $\psi(a,y^{1/g}) = \kappa(a)$ and 
$\psi(a,y) = \kappa(a)$.  This shows that
$\liminf_{n\to\infty} \Sfrac{1}{fn} \log S_{fn}^{(f)} (a,y) \geq \kappa(a)$.

Finally, suppose $a>a_c$ and $y>1$. 

If $\kappa(a)>\lambda(y)$ then the right hand side is a maximum if $g=f-1$
so that $\liminf_{n\to\infty} \Sfrac{1}{fn} \log S_{fn}^{(f)} (a,y) \geq \kappa(a)$.

On the other hand, if $\kappa(a) < \lambda(y^{1/(f-1)}) < \lambda(y)$ then
$\psi(a,y^{1/g}) = \lambda(y^{1/g})$, and by equation \Ref{eqn26},
$\liminf_{n\to\infty} \Sfrac{1}{fn} \log S_{fn}^{(f)} (a,y) \geq
\max\{ \sfrac{g}{f}\, \lambda(y^{1/g}) + \sfrac{1}{f}\, \lambda(y)
+ \sfrac{f-g-1}{f} \, \log\mu_3 \} = \sfrac{2}{f} \, \lambda(y) + \sfrac{f-2}{f} \, \log \mu_3$.

The only remaining case is when there exists an $h$ such that
\[ \lambda(y^{1/(f-1)}) \leq \ldots \leq \lambda(y^{1/h})
< \kappa(a) \leq \lambda(y^{1/(h-1)}) \leq \ldots \leq \lambda(y) . \]
In this case for $h \leq g < f$, $\psi(a,y^{1/g}) = \kappa(a)$, and
for $1\leq g < h$, $\psi(a,y^{1/g}) = \lambda(y^{1/g})$.  In other words,
\begin{equation}
\liminf_{n\to\infty} \Sfrac{1}{fn} \log S_{fn}^{(f)} (a,y) \geq \max
\cases{
\max_{1\leq g < h} \LC  \Sfrac{g}{f}\, \lambda(y^{1/g}) + \Sfrac{1}{f} \lambda(y) + \Sfrac{f-g-1}{f} \, \log\mu_3 \RC & \\
\max_{h\leq g < f} \LC  \Sfrac{g}{f}\, \kappa(a) + \Sfrac{1}{f} \lambda(y) + \Sfrac{f-g-1}{f} \, \log\mu_3 \RC &
} .
\end{equation}
The maximum in the first bound is clearly when $g=1$ (by equation \Ref{eqn26}),
and in the second bound, when $g=f-1$.  That is, when
\begin{equation}
\liminf_{n\to\infty} \Sfrac{1}{fn} \log S_{fn}^{(f)} (a,y) \geq \max
\cases{
\Sfrac{2}{f} \lambda(y) + \Sfrac{f-2}{f} \, \log\mu_3  & \\
\Sfrac{f-1}{f}\, \kappa(a) + \Sfrac{1}{f} \lambda(y)  &
} .
\end{equation}
Putting these results together gives the following lemma:

\begin{lemm}
$\displaystyle
\liminf_{n\to\infty} \Sfrac{1}{fn} \log  {S}^{(f)}_{fn} (a,y) \geq
\max\LC 
\kappa(a), 
\Sfrac{f-1}{f}\, \kappa(a) + \Sfrac{1}{f} \, \lambda(y),
\Sfrac{2}{f} \, \lambda(y) + \Sfrac{f-2}{f} \, \log \mu_3 \RC $. \qed
\end{lemm}

An upper bound equal to this lower bound is obtained by considering the 
spider and the arms of the star to be independent.  

If the pulled star is composed of just a pulled arm, and an adsorbing spider,
then suppose the height of the pulled vertex is $h$ above the adsorbing plane, 
and the height of the central node of the spider is $h_1$ (so that the vertical
extent of the pulled arm is $h-h_1$). Combining the pulled arm
and the spider is a convolution of the partition functions of the two objects, which
becomes a sum of the free energies in the thermodynamic limit (after taking
logarithms).  This shows that the free energy contribution of a star which is
composed of just a pulled arm and a spider (with $g$ legs) is 
$\sfrac{g}{k} \, \psi(a,y^{1/g}) + \sfrac{1}{k}\, \psi(a,y)$ (where $k=g+1$).

If there are $f>k$ arms in the star, with one pulled, $g$ in an adsorbing spider,
and $f-g-1$ remaining arms are disjoint with the adsorbing plane,  then the
free energy is bounded above by
\begin{equation}
\limsup_{n\to\infty} \Sfrac{1}{fn} \log S_{fn}^{(f)} (a,y) \leq
\max_{g\geq 1} \L \sfrac{g}{f} \, \psi(a,y^{1/g}) + \sfrac{1}{f}\, \psi(a,y) 
+ \Sfrac{f-g-1}{f} \, \log \mu_3 \R .
\end{equation}
Notice that this is the maximum of the same expression obtained in 
equation \Ref{eqn29}, and this was analysed above.  The result is that

\begin{lemm}
$\displaystyle
\limsup_{n\to\infty} \Sfrac{1}{fn} \log  {S}^{(f)}_{fn} (a,y) \leq
 \max\LC 
\kappa(a), 
\Sfrac{f-1}{f}\, \kappa(a) + \Sfrac{1}{f} \, \lambda(y),
\Sfrac{2}{f} \, \lambda(y) + \Sfrac{f-2}{f} \, \log \mu_3 \RC $. \qed
\end{lemm}

Putting the last two lemmas together gives the following theorem.

\begin{theo}
The free energy of pulled adsorbing uniform $f$-stars pulled at the end-vertex
of an arm is given by
$$
\sigma^{(f)} (a,y) = 
\lim_{n\to\infty} \Sfrac{1}{fn} \log  {S}^{(f)}_{fn} (a,y)
= \max\LC 
\kappa(a), 
\Sfrac{f-1}{f}\, \kappa(a) + \Sfrac{1}{f} \, \lambda(y),
\Sfrac{2}{f} \, \lambda(y) + \Sfrac{f-2}{f} \, \log \mu_3 \RC .
\; \square $$
\end{theo}

The locations of the phase boundaries are given by solutions of 
\begin{equation}
\lambda(y)=\kappa(a),\quad\hbox{and}\quad
\lambda(y) = (f-1)\kappa(a) - (f-2)\log \mu_3. 
\end{equation}
Notice that $f>1$ in this model.  These phase boundaries separate 
an adsorbed phase with $\sigma^{(f)} (a,y)= \kappa(a)$, a mixed
phase with $\sigma^{(f)}(a,y) = \sfrac{f-1}{f}\, \kappa(a) + \sfrac{1}{f} \, \lambda(y)$
and a ballistic phase with $\sigma^{(f)} (a,y)= \sfrac{2}{f} \, \lambda(y) + \sfrac{f-2}{f} \, \log \mu_3$.
These phase boundaries are first order transitions. 

If $f=2$, then $\sigma^{(2)} (a,y)= \max\{\kappa(a), \sfrac{1}{2}(\kappa(a)+\lambda(y)),
\lambda(y)\} = \max\{\kappa(a),\lambda(y)\} = \psi(a,y)$.  This is expected, since
in this case the model is an adsorbing self-avoiding walk of length $2n$ pulled 
at its free endpoint \cite{Rensburg2016b}.

If $f=3$, then
\begin{equation}
\sigma^{(3)}(a,y)= \max\{\kappa(a), \sfrac{1}{3}(2\kappa(a)+\lambda(y)),
\sfrac{1}{3}(2\lambda(y)+\log \mu_3)\}.
\end{equation}
This is the free energy of an adsorbing $3$-star pulled at a vertex of degree
$1$ (see reference \cite{Rensburg2018}).  The phase diagram of this model
has in addition to the free, ballistic and adsorbed phases, a mixed phase which
is both ballistic and adsorbed (corresponding to the case that
$\sigma^{(3)}(a,y) = \sfrac{1}{3}(2\kappa(a)+\lambda(y))$ above).  An adsorbed
$3$-star pulled at a vertex of degree $1$ will first desorb the pulled arm which
becomes ballistic, then desorb the remaining adsorbed arm when the attached
arm becomes ballistic.

If $f=4$, then the free energy is given by
\begin{equation}
\sigma^{(4)}(a,y)= \max\{\kappa(a), \sfrac{1}{4}(3\kappa(a)+\lambda(y)),
\sfrac{1}{4}(2\lambda(y)+2\log \mu_3)\}.
\end{equation}
Similar to the case $f=3$,  there is a free phase with $\sigma^{(4)}(a,y)=\log \mu_3$,
an adsorbed phase with $\sigma^{(4)} (a,y) = \kappa(a)$ (all four arms are adsorbed),
a ballistic phase with $\sigma^{(4)} (a,y) = \sfrac{1}{4} (2\lambda(y) + 2\log \mu_3)$
(two arms are ballistic and two are free), and a mixed phase with
$\sigma^{(4)} (a,y) = \sfrac{1}{4} (3\kappa(a) + \lambda(y))$ (three arms adsorbed
and the pulled arm ballistic).  In other words, an adsorbed $4$-star pulled by
a vertex of degree $1$ will first desorb the pulled arm (which becomes ballistic),
then desorb two more arms simultaneously (which become free) when the attached
arm also becomes ballistic.  This generalizes to $f>4$ by $f-2$ arms desorbing 
simultaneously when the attached arms become ballistic.

\section{Uniform tadpoles and related graphs}
\label{sec:tadpoles}

We first consider \emph{tadpoles} (see figure \ref{fig:models}(b) and \ref{fig:models}(c)).  
These are connected graphs with cyclomatic index 
1, one vertex of degree 3 and one vertex of degree 1.  A tadpole is uniform if the number of edges in 
the circuit (often called the \emph{head} of the tadpole) is equal to the number of 
edges between the vertex of degree 1 and the vertex of degree 3 (often called
the \emph{tail}).  We write $t_n$ for the number of embeddings of uniform tadpoles with $n$ edges
in $\mathbb{Z}^3$, so that there are $\sfrac{1}{2}n$ edges in the head and $\sfrac{1}{2}n$ edges
in the tail.  The head must have an even number of edges (since the cubic lattice is bipartite) so 
$n$ must be divisible by 4.  It is known that 
\cite{Soteros,Soteros1}:
\begin{equation}
\lim_{n\to\infty} \Sfrac{1}{n}  \log t_n = \log \mu_3.
\end{equation}

There is a particular vertex in the head, connected to the vertex of degree 3 by two 
walks each with $\sfrac{1}{4}n$ edges (since the head has length $\sfrac{1}{2}n$ edges).  We call
this the \emph{middle vertex}.  Consider uniform tadpoles with their middle vertex at the origin
(see figure 1(b)), 
confined to $x_3 \ge 0$ and subject to a force at the vertex of degree 1.  We shall need a preliminary lemma.

\begin{lemm}
If a polygon in $\mathbb{Z}^3$ has one  vertex attached at the origin and is pulled 
from the middle vertex the free energy is given by $\max\{\kappa(a),\lambda(\sqrt{y})\}$.
\label{lem:polygon}
\end{lemm}
\Pr By Soteros 1992 \cite{Soteros} the free energy of adsorbing polygons
in $\mathbb{Z}^3$ is $\kappa(a)$, and by theorems 5 and 6 in reference \cite{Guttmann2018}
the free energy of a polygon pulled in its midpoint is $\lambda(\sqrt{y})$.
Thus, the free energy is bounded above by $\max\{ \kappa(a),\lambda(\sqrt{y}) \}$.
This is also a lower bound, since the free energy of a pulled adsorbing polygon cannot
be less than $\kappa(a)$, or less that $\lambda(\sqrt{y})$.
It follows that the free energy of a pulled attached adsorbing polygon is the maximum
of these free energies.
\qed

Let $t_n^{(1)}(v,h)$ be the number of uniform tadpoles with a total of $n$ 
edges, confined to $x_3 \ge 0$, with $v+2$ vertices in the plane
$x_3=0$, having their middle 
vertex at the origin and with the $x_3$ coordinate of the vertex of degree 1 equal
to $h$.
Our primary result in this section is the following theorem, covering the case where $a \ge1$
and $y \ge 1$, so that the tadpole is attracted to the surface but with the force
directed away from the surface.

\begin{theo}
Suppose that $a \ge 1$ and $y \ge 1$.
For uniform tadpoles with their middle vertex at the origin and pulled at the 
vertex of degree 1, the free energy is given by
$$\tau^{(1)}(a,y)=\lim_{n\to\infty} \Sfrac{1}{n}  \log \sum_{v,h} t_n^{(1)}(v,h) a^v y^h 
= \max\left\{\kappa(a), \Sfrac{1}{2} (\kappa(a) + \lambda(y)), 
\Sfrac{1}{2} (\lambda(y) + \lambda(\sqrt{y})) \right\}.$$
\end{theo}
\Pr
The proof proceeds by establishing upper and lower bounds.   We first consider the lower bounds.
Soteros \cite{Soteros} has proved that 
\begin{equation}
\tau^{(1)}(a,1)=\lim_{n\to\infty} \Sfrac{1}{n}  \log \sum_{v,h} t_n^{(1)}(v,h) a^v = \kappa(a).
\end{equation}
Since $\sum_{v,h} t_n^{(1)}(v,h)a^vy^h$ is monotone increasing in $y$ we have
\begin{equation}
\liminf_{n\to\infty} \Sfrac{1}{n}  \log \sum_{v,h} t_n^{(1)} (v,h) a^vy^h \ge \kappa(a)
\end{equation}
when $y \ge 1$.
Consider a polygon with $\sfrac{1}{2}n$ edges rooted at the origin and with the middle vertex 
in the top plane.  Attached at the middle vertex is a bridge with $\sfrac{1}{2}n$ edges.  This
bridge has no vertices in $x_3=0$, by construction.  The free energy is the sum of the 
free energy contributions from the polygon and the bridge so 
\begin{equation}
\liminf_{n\to\infty} \Sfrac{1}{n}  \log \sum_{v,h}  t_n^{(1)}(v,h) a^v y^h  \ge
\Sfrac{1}{2} \left( \max\left\{\kappa(a),\lambda(\sqrt{y})\right\}+\lambda(y)\right) .
\end{equation}
We now derive corresponding upper bounds by considering the head and tail to be
independent.  The head of the tadpole certainly has
vertices in $x_3=0$.  Suppose first that the tail has no vertices in $x_3=0$.  The contribution
from the head is at most $\sfrac{1}{2}\max[\kappa(a),\lambda(\sqrt{y})]$ and the contribution from the tail 
is $\sfrac{1}{2}\lambda(y)$ since the tail cannot visit $x_3=0$.  Now suppose that the tail does have vertices
in $x_3=0$.  Then the head is not under tension and contributes $\sfrac{1}{2}\kappa(a)$ while the tail
contributes at most $\sfrac{1}{2}\max[\kappa(a),\lambda(y)]$.  These upper bounds show that
\begin{equation}
\hspace{-0.5cm}
\limsup_{n\to\infty} \Sfrac{1}{n} 
 \log \sum_{v,h} t_n^{(1)}(v,h) a^vy^h \le \max\left\{\kappa(a),\Sfrac{1}{2}(\kappa(a)+\lambda(y)),
\Sfrac{1}{2}(\lambda(\sqrt{y})+\lambda(y))  \right\}.
\end{equation}
These upper bounds match the lower bounds given above and complete the proof.
\qed

\begin{theo}
The free energy of
uniform tadpoles, with their middle vertex at the origin and pulled at the 
vertex of degree 1,
when $a \le 1$ is
$$\tau^{(1)}(a,y) = \Sfrac{1}{2}(\lambda(y)+\lambda(\sqrt{y})),$$
and the free energy when $y \le 1$ is
$$\tau^{(1)}(a,y) = \kappa(a).$$
In particular, when $y \le 1$ and $a \le a_c$ the free energy is $\log \mu_3$.
\label{theo:t1mu}
\end{theo}
\Pr 
We first note that a polygon in $\mathbb{Z}^3$ with vertices at $(0,0,0)$ 
and $(1,0,0)$ and with the edge
joining these two vertices, confined to the octant $x_1 \ge 0$, $x_2 \ge 0$, $x_3 \ge 0$,
and interacting with the plane $x_3=0$ has free energy $\kappa(a)$.  This can be proved by the 
methods developed in \cite{HammersleyWhittington} and \cite{Soteros}.  

We first deal with the 
case $y \le 1$.  We have the obvious upper bound (that the free energy is bounded
above by $\kappa(a)$) from monotonicity in $y$ so we only need
a lower bound.   We construct a subset of tadpoles with middle vertex at the origin and the 
vertex of degree 1 in $x_3=0$.  We first construct two polygons, $A$ and $B$,
each with $\sfrac{1}{4}n - 1$ edges.  Polygon $A$ has a vertex at $(0,1,0)$ and an adjacent vertex 
at $(1,1,0)$, and is confined to the octant $x_1 \ge 0$, $x_2 \ge 1$, $x_3 \ge 0$.
Polygon $B$ has a vertex at $(0,-1,0)$ and an adjacent vertex at $(1,-1,0)$, and
is confined to the octant $x_1 \ge 0$, $x_2 \le -1$, $x_3 \ge 0$.   We concatenate 
$A$ and $B$ to form a polygon, $C$ by deleting the edges $(0,1,0)-(1,1,0)$
and $(0,-1,0)-(1,-1,0)$ and adding four edges $(0,1,0)-(0,0,0)$, $(0,0,0)-(0,-1,0)$,
$(1,1,0)-(1,0,0)$ and $((1,0,0)-(1,-1,0)$.  This polygon has $\sfrac{1}{2}n$ edges, a vertex at 
$(0,0,0)$ and a vertex at $(1,0,0)$.  These two vertices are opposite since they are
separated by two walks each with $\sfrac{1}{4}n$ edges.  Next add the edge $(0,0,0)-(-1,0,0)$
and a loop with $\sfrac{1}{2}n-1$ edges, unfolded in the $x_1$-direction so 
that it has no vertices with $x_1 > -1$
and the resulting graph is a tadpole with the vertex of degree 3 at the origin.  
These graphs have free energy $\kappa(a)$ and are a subset of the tadpoles 
contributing to $\sum_{v,h} t_n^{(1)}a^vy^h$ for all $y$.   This completes the proof that the free 
energy is equal to $\kappa(a)$ for all $y \le 1$.  

In order to deal with the situation when 
$a \le 1$ we again note that the free energy is $\sfrac{1}{2}(\lambda(y)+\lambda(\sqrt{y}))$ 
when $a=1$ so this provides an upper bound.
To construct a lower bound, consider the subset of tadpoles such that the
polygon (\emph{i.e.} the head) has exactly two
vertices in $x_3=0$.  These are an appropriate subset for all $a \le 1$.
Every polygon can be converted to such a polygon
by adding two edges and translating an edge so it is easy to see that this subset of 
tadpoles has the same free energy as when $a=1$.  This completes the proof.
\qed

There are four phases, a free phase (if $y < 1$ and $a < a_c$), a ballistic phase
(if $y>1$ and $\lambda(\sqrt{y}) > \kappa(a)$), an adsorbed phase (if $a > a_c$ 
and $\kappa(a) > \lambda(y)$) and
a mixed phase.  In the mixed phase the head of the tadpole is adsorbed but the tail 
is ballistic.  The applied force is large enough to desorb the tail but not large 
enough to desorb the head.

Alternatively we can fix the vertex of degree 1 at the origin and pull from the middle
vertex.  
Let 
$t_n^{(2)}(v,h)$ be the number of uniform tadpoles with a total of $n$ 
edges, confined to $x_3 \ge 0$, with $v+1$ vertices in the plane
$x_3=0$, having the vertex of degree 1 
at the origin.  Suppose that the vertex of the polygon (forming the head) that is 
$\sfrac{1}{4}n$ edges from the vertex of degree 3 has $x_3$ coordinate equal
to $h$.
Then we have the following theorem:

\begin{theo}
Suppose that $a \ge 1$ and $y \ge 1$.  For uniform tadpoles with their vertex of degree 1 at the origin and pulled at the middle
vertex of the head, the free energy is given by
$$\tau^{(2)}(a,y) = \lim_{n\to\infty} \Sfrac{1}{n}  \log \sum_{v,h} t_n^{(2)}(v,h)a^vy^h
=\max\left\{ \kappa(a),  \Sfrac{1}{2}(\lambda(y)+\lambda(\sqrt{y})) 
\right\}.$$
\end{theo}
\Pr
We obtain one lower bound by noting that 
$\tau^{(2)}(a,1) = \kappa(a)$ \cite{Soteros} and that $\sum_{v,h} t_n^{(2)}(v,h)a^vy^h$
is monotone increasing in $y$ for $y > 1$.  This implies that
$$\liminf_{n\to\infty} \Sfrac{1}{n}  \log \sum_{v,h} t_n^{(2)}(v,h) a^vy^h \ge \kappa(a).$$
If we set $a=1$ and pull on a tadpole made up of a bridge with $\sfrac{1}{2}n$ edges 
concatenated with a polygon so that the degree 3 vertex is in the 
bottom plane of the polygon and the opposite vertex is in the top plane,
then, using the methods of reference \cite{Guttmann2018}, it can be shown that
$$\tau^{(2)}(1,y) = \Sfrac{1}{2}(\lambda(y)+\lambda(\sqrt{y}) ).$$
Since $\sum_{v,h} t_n^{(2)}(v,h) a^vy^h$ is monotone non-decreasing  in $a$ we 
have
$$\liminf_{n\to\infty} \Sfrac{1}{n}  \log \sum_{v,h} t_n^{(2)}(v,h) a^vy^h \ge \Sfrac{1}{2}(\lambda(y)+\lambda(\sqrt{y}) ).$$
We can construct upper bounds by treating the head and tail as being independent.  If the head does not have vertices in $x_3=0$ the contribution to the free energy is at most
$$\Sfrac{1}{2}\left(\max\{\kappa(a),\lambda(y)\}+\lambda(\sqrt{y}) \right),$$
while if the head has vertices in $x_3=0$ the maximum contribution to the free energy is
$$\Sfrac{1}{2}\left(\kappa(a)+\max\{\kappa(a),\lambda(\sqrt{y})\}\right).$$
Finally we note that, when $y \ge 1$, $ \lambda(\sqrt{y}) > \kappa(a)$ implies
that $\lambda(y) > \kappa(a)$.  Hence 
\begin{equation}
\limsup_{n\to\infty} \Sfrac{1}{n}  \log \sum_{v,h} t_n^{(2)}(v,h) a^vy^h \le 
\max\left\{ \kappa(a), \Sfrac{1}{2}(\lambda(y)+\lambda(\sqrt{y}) ) \right\}
\end{equation}
and the upper and lower bounds taken together complete the proof.
\qed

\begin{theo}
The free energy of
uniform tadpoles, with their vertex of degree 1 at the origin and pulled at the middle
vertex of the head,
when $a \le 1$ is
$$\tau^{(2)}(a,y) = \Sfrac{1}{2} ( \lambda(y)+\lambda(\sqrt{y}) ),$$
and the free energy when $y \le 1$ is
$$\tau^{(2)}(a,y) = \kappa(a).$$
In particular, when $y \le 1$ and $a \le a_c$ the free energy is $\log \mu_3$.

\label{theo:t2mu}
\end{theo}
\Pr 
For the case when $y \le 1$ the proof is essentially identical to the proof of 
Theorem \ref{theo:t1mu} and the tadpole used for the lower bound is identical 
to the one constructed in that proof.
This deals with the situation when $a \le 1$ and $y \le 1$ so we need to consider 
$a \le 1$ and $y > 1$.  We construct a lower bound as follows.  Begin with the edge 
$(0,0,0)-(0,0,1)$ and add to this a walk with $\sfrac{1}{2}n-2$ edges, unfolded in the $x_3$-direction, 
followed by an additional edge in the $x_3$-direction.  This walk has exactly
one vertex in $x_3=0$.  Consider a polygon with $\sfrac{1}{2}n$ edges with a vertex in the bottom 
plane opposite to a vertex in the top plane.  Concatenate this polygon and the 
walk already constructed at the distinguished vertex in the bottom plane,
and apply a force at the opposite vertex.  The free energy is 
$\sfrac{1}{2}(\lambda(y)+\lambda(\sqrt{y}))$.  Since these polygons have only one vertex in 
$x_3=0$ they are a lower bound for polygons when $y > 1$ and $a > 0$.  This completes the 
proof.  Since $\kappa(a) = \log \mu_3$ when $a \le a_c$ and $\lambda(y) =
\lambda(\sqrt{y})=\log \mu_3$ when $y \le 1$ the free energy is 
$\log \mu_3$ when $a \le a_c$ and $y \le 1$.
\qed

In this case there is no mixed phase.

We can treat dumbbells (see figure 1(d)) similarly.  These are connected graphs with cyclomatic 
index 2 and two vertices of degree 3.   They can be thought of as two polygons joined by a 
walk.  Since dumbbells have two cycles 
they have two middle vertices, one in each cycle.  Let $d_n(v,h)$ be the number of 
uniform dumbbells confined to $x_3 \ge 0$ with a middle vertex at the origin and the other middle
vertex at height $h$ above $x_3=0$, and $v+2$ vertices in $x_3=0$.  Define the partition function 
\begin{equation}
D_n(a,y) = \sum_{v,h} d_n(v,h) a^v y^h.
\end{equation}

We have the following theorem.

\begin{theo}
Let $a \ge 1$ and $y \ge 1$.  For uniform dumbbells with one middle vertex at the origin and pulled at the other middle
vertex, the free energy is given by
$$ \lim_{n\to\infty} \Sfrac{1}{n} \log D_n(a,y) = \max\left\{ \kappa(a),  
\Sfrac{1}{3}(2\lambda(\sqrt{y})+\lambda(y) ) \right\}.$$
\end{theo}
\Pr
We first construct a  lower bound by considering a subset of 
dumbbells.  If $y=1$ the free energy is $\kappa(a)$ \cite{Soteros} so $\kappa(a)$
is a lower bound for all $y \ge 1$.  Now suppose that $a=1$ 
and consider a polygon with $\sfrac{1}{3}n$ edges with a vertex at the origin, confined to $x_3 \ge 0$, with its middle vertex in the top plane of 
the polygon.  Attach an edge at this middle vertex in the positive $x_3$-direction, add a 
walk with $\sfrac{1}{3}n-2$ edges, unfolded in the $x_3$-direction, an additional edge
in this direction and then a polygon with $\sfrac{1}{3}n$ edges with its bottom vertex 
a attached to this edge and its middle vertex in the top plane of the polygon.  The free 
energy is $\sfrac{1}{3}(2\lambda(\sqrt{y}) + \lambda(y))$ and this is a lower bound for all
$a \ge 1$.  To get an upper bound we treat the two polygons and the walk as being 
independent.  We need to consider three cases.  Suppose that only the first polygon
(the one with a vertex in $x_3=0$) has vertices in $x_3=0$.  Then the free energy contribution is 
at most
$$\Sfrac{1}{3}\left( \max\{\kappa(a),\lambda(\sqrt{y}) \} + \lambda(y) + \lambda(\sqrt{y}) \right).$$
If the walk has vertices in $x_3=0$ the first polygon is not
subject to a force and the total free energy contribution is at most
$$\Sfrac{1}{3} \left( \kappa(a) + \max\{\kappa(a), \lambda(y)\}+\lambda(\sqrt{y})  \right).$$
Finally, if the second polygon (where the force is applied) has vertices 
in $x_3=0$ the free energy contribution is at most
$$\Sfrac{1}{3} \left( 2\kappa(a) + \max\{\kappa(a), \lambda(\sqrt{y})\} \right).$$
Recall that when $y > 1$ $\lambda(y) > \lambda(\sqrt{y})$.  Of these six possibilities
the maximum is either $\kappa(a)$ or 
$\sfrac{1}{3}(2\lambda(\sqrt{y})+\lambda(y))$.  Together with
the two lower bounds above this completes the proof.
\qed

It remains to consider the situation when either $y < 1$ or $a < 1$.  
\begin{theo}
When $y \le 1$ the free energy of dumbbells is $\kappa(a)$ and when 
$a \le 1$ the free energy is $\sfrac{1}{3}(2 \lambda(\sqrt{y})+\lambda(y))$.  In particular,
when $y \le 1$ and $a \le a_c$ the free energy is $ \log \mu_3$.
\end{theo}
\Pr We omit the proof of this theorem since it is very similar to the
proof of theorem \ref{theo:t2mu}. \qed

There is no mixed phase.  For $y > 1$, $\lambda(y) > \lambda(\sqrt{y})$ so if the force 
is large enough to pull the first circuit off the surface then it is large enough to pull
off both the walk between the two vertices of degree 3 and the other circuit.  Hence there
is a free phase (when $y<1$ and $a<a_c$), an adsorbed phase (when $a> a_c$ and 
$\kappa(a) > \sfrac{1}{2} (2\lambda(\sqrt{y})+\lambda(y))$) and a ballistic phase (when $y > 1$ and 
$\kappa(a) < \sfrac{1}{2} (2\lambda(\sqrt{y})+\lambda(y))$).

\begin{figure}[t]
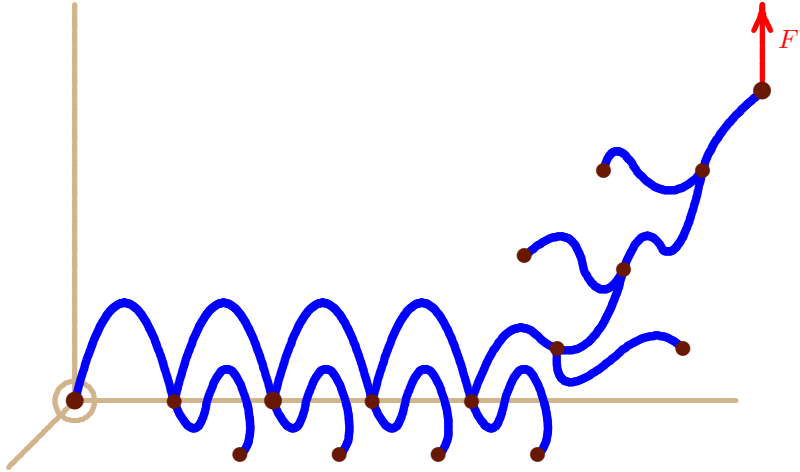

\beginpicture
\setcoordinatesystem units <2.5pt,2.5pt>
\setplotarea x from -40 to 160, y from -10 to 65
\setplotarea x from 0 to 150, y from -10 to 60

\color{Tan}
\setplotsymbol ({\LARGE$\cdot$})

\setsolid
\plot 0 60 0 0 100 0 /
\plot 0 0 -10 -10 /

\circulararc 360 degrees from 0 3 center at 0 0 

\color{black}

\color{red}
\arrow <10pt> [.2,.67] from 104 47 to 104 60
\put {$F$} at 108 55

\setplotsymbol ({\scriptsize$\bullet$})
\color{Blue}

\multiput 
{\beginpicture
\setquadratic
\setplotsymbol ({\scriptsize$\bullet$})
\color{Blue}
\plot  0 0 7.5 15 15 0  /
\plot 15 0 18 -4 20 0 23 5 26 0  26.5 -5 25 -8 /
\setlinear
\setplotsymbol ({\LARGE$\cdot$})
\color{Sepia}
\put {\Large$\bullet$} at 0 0 
\endpicture}
at  0 0 15 0 30 0 45 0   /
\color{Sepia}
\put {\LARGE$\bullet$} at 0 0 
\multiput {\Large$\bullet$} at 25 -8 40 -8 55 -8 70 -8 /

\setquadratic
\color{Blue}
\plot 60 0 65 10 70 10 73 8 76 8 80 12 83 20 86 25 89 23 92 25 95 35 98 41 104 47  /
\plot 73 8  75 3 83 8 88 10 92 8 /
\plot 83 20 80 17 77 20 74 25 68 22 /
\plot 95 35 90 32 85 35 82 38 80 35 /
\color{Sepia}
\multiput {\Large$\bullet$} at 60 0 73 8 83 20 95 35 104 47 92 8 68 22 80 35  /

\setlinear
\color{Sepia}
\multiput {\LARGE$\bullet$} at 30 0  104 47  /

\color{black}
\normalcolor
\endpicture
\caption{Pulling an adsorbed comb at its endpoint.  In this drawing three teeth are desorbed and the last
vertex in the adsorbing plane is the junction of an (adsorbed) tooth and the backbone of the comb.
There are four adsorbed teeth.}
\label{figure10} 
 \end{figure}

\section{Uniform combs}
\label{sec:combs}

In this section we examine uniform combs.  See Figure 1 for a sketch.  A comb can be 
thought of as a self-avoiding walk making up the backbone of the comb, with $t$ 
teeth attached at regular intervals along the backbone.  There are $t$ vertices of degree 3
and each branch, either between two vertices of degree 3 or 
between a vertex of degree 3 and a vertex 
of degree 1, has the same number of edges.

We consider the case $a>a_c$ and $y>1$ first.

Suppose that each branch in the comb has length $m$.  
Then the total size of the comb is $N=(2t+1)m$ edges if
there are $t$ teeth.  Let the number of combs with one endpoint at the origin,
with $t$ teeth, of total size (number of edges) $N$, making $v$ visits to the adsorbing plane 
and other endpoint at height $h$, be $k_N^{(t)}(v,h)$.  The partition function is given by
\begin{equation}
K_N^{(t)}(a,y) = \sum_{v,h} k_N^{(t)}(v,h)\, a^vy^h .
\end{equation}

We outline a proof of a lower bound 
on $\liminf_{m\to\infty} \sfrac{1}{(2t+1)m}\log K_{(2t+1)m}^{(t)}(a,y)$.

Consider a partially desorbed comb with $t$ teeth as illustrated schematically in figure \ref{figure10}.
The loops along the adsorbed part of the comb are doubly unfolded with endpoints in the adsorbing
plane, and the associated adsorbed teeth are quarantined in $\alpha$-wedges for a small angle $\alpha$.
The desorbed part of the comb is a doubly unfolded positive walk, and the teeth along it are also desorbed
and quarantined to $\alpha$-wedges.

The first part of the backbone is adsorbed, and the remaining part is pulled off by the force $F$.   
The part of the backbone pulled from the adsorbing surface has length $(s+1)m$, so that $s$
teeth are desorbed.  The length of the adsorbed part is $(t-s)m$ and so there are
$t-s$ teeth that are adsorbed. Thus, each adsorbed tooth contributes
$\kappa(a)$ to the free energy, and the adsorbed backbone contributes $(t-s)\,\kappa(a)$.  The first
part of the backbone following the last adsorbed tooth contributes $\psi(a,y)$, the remaining
part of the backbone pulled off contributes $s \lambda(y)$, and the remaining teeth on this backbone $s \log \mu_3$.  
This shows that 
\[ \hspace{-2cm}
\liminf_{m\to\infty} \sfrac{1}{(2t+1)m}\log K_{(2t+1)m}^{(t)}(a,y)
 \geq \Sfrac{1}{2t+1} (2(t-s)\,\kappa(a)+s\,\lambda(y)+\psi(a,y)+ s\,\log \mu_3) .\]
Since $s$ is arbitrary, take the maximum on the right hand side to find the best lower bound.

Rearrange terms in the last expression as follows:
\begin{equation}  \hspace{-2cm}
\liminf_{m\to\infty} \sfrac{1}{(2t+1)m}\log K_{(2t+1)m}^{(t)}(a,y)
 \geq \Sfrac{1}{2t+1} (2t\,\kappa(a)+\psi(a,y) + s (\lambda(y)+\log \mu_3 - 2\,\kappa(a)) .
\label{eqn48}   
\end{equation}
If $\kappa(a) > \sfrac{1}{2}( \lambda(y) + \log \mu_3)$, then the last term 
is negative and the best lower bound is obtained by putting $s=0$ so
that the free energy is bounded below by  $\sfrac{1}{2t+1} (2t\,\kappa(a)+\psi(a,y))$.
This gives two possibilities.  

Either $\sfrac{1}{2}( \lambda(y) + \log \mu_3) < \kappa(a) < \lambda(y)$
in which case $\psi(a,y) = \lambda(y)$, or $\lambda(y) < \kappa(a)$ 
so that $\psi(a,y)=\kappa(a)$.  Taken together, this gives the lower bound
\[  \hspace{-2.5cm}
\liminf_{m\to\infty} \sfrac{1}{(2t+1)m}\log K_{(2t+1)m}^{(t)}(a,y)
 \geq  \max\{ \kappa(a),\sfrac{2t}{2t+1}\,\kappa(a)+\sfrac{1}{2t+1}\,\lambda(y) \},
\quad\hbox{if $\kappa(a) > \sfrac{1}{2}( \lambda(y) + \log \mu_3)$}.
\]
On the other hand, if $\kappa(a) < \sfrac{1}{2}( \lambda(y) + \log \mu_3)$, then the best lower bound
is obtained if $s=t$ in equation \Ref{eqn48}.  This shows that the free energy is bounded
below by  $\sfrac{1}{2t+1} (t\,\lambda(y) + \psi(a,y) + t\log \mu_3)
= \sfrac{1}{2t+1}((t+1)\,\lambda(y) + t \log \mu_3)$ since $\psi(a,y)=\lambda(y)$
because  $\kappa(a) < \sfrac{1}{2}( \lambda(y) + \log \mu_3) < \lambda(y)$.  
Combining this with the last bound above gives the following lemma:

\begin{lemm}
If $a>a_c$ and $y>1$, then
$$\displaystyle
\liminf_{m\to\infty} \sfrac{1}{Tm}\log K_{Tm}^{(t)}(a,y)
 \geq \max\{ \kappa(a),\sfrac{2t}{2t+1}\,\kappa(a)+\sfrac{1}{2t+1}\,\lambda(y),
\sfrac{t+1}{2t+1}\,\lambda(y) + \sfrac{t}{2t+1}\log \mu_3 \} ,
$$
 where $T=2t+1$. \qed
\label{lemma6}
\end{lemm}

To find upper bounds, we first establish some notation.  Consider a comb with 
$t$ teeth and number the teeth $j=1,2, \ldots t$.  Corresponding to each tooth 
is a degree $3$ vertex with the same label.  There are $t+1$ branches in the 
backbone of the comb and we number these $j=0,1,2, \ldots t$ with the branch 
labelled zero attached at the origin and with the force applied at the degree $1$ 
vertex in the branch labelled $t$.  The first case to consider is when
the comb is adsorbed and the free energy is $\kappa(a)$.  Otherwise
suppose that all vertices after the branch point labelled $s$ are desorbed and 
that there are adsorbed vertices either in the tooth labelled $s$ or in the backbone
branch labelled $s$, or in both.    To get an upper bound we subdivide the comb into
three parts: 
\begin{enumerate}
\item
the branch of the backbone labelled $s-1$ and the tooth labelled $s$, 
\item
the part of the comb up to the vertex of degree 3 labelled $s-1$, including the $s-1$ 
tooth, and
\item
the sub-comb consisting of everything after the degree 3 vertex 
labelled $s$,
\end{enumerate} 
and regard the three parts as independent.  Part (ii) has $2(s-1)$ branches and contributes
$\sfrac{2(s-1)}{2t+1} \kappa(a)$ to the free energy.  Part (iii) is a comb under tension 
and contributes $\sfrac{t+1-s}{2t+1}\lambda(y) + \sfrac{t-s}{2t+1} \log \mu_3$ to the free 
energy.  For part (i) we have a contribution of 
$$\Sfrac{1}{2t+1}\max\{2\kappa(a),2\lambda(\sqrt{y}),\lambda(y)+\log \mu_3\}$$
since both branches can be adsorbed or the two branches can be pulled as a loop (if the 
tooth is adsorbed) or as a backbone branch plus a free tooth.  But 
$2 \lambda(\sqrt{y}) \le \lambda(y) + \log \mu_3$ by convexity.  If $2 \kappa(a) >
\lambda(y) + \log \mu_3$ the expression for the free energy is maximized when 
$s=t$ and if $\lambda(y) + \log \mu_3 > 2\kappa(a)$ it is maximized when $s=1$.
Finally, combining this with the bound for the adsorbed comb, this yields the upper bound
\begin{equation}
\hspace{-2cm}
\limsup_{m\to\infty} \sfrac{1}{Tm}\log K_{Tm}^{(t)}(a,y) \leq
\max\{\kappa(a), \sfrac{t+1}{2t+1}\lambda(y)+\sfrac{t}{2t+1}\log \mu_3, 
\sfrac{2t}{2t+1} \kappa(a) + \sfrac{1}{2t+1}\lambda(y)] \},
\end{equation}
where $T=2t+1$.
Comparing this to the bound in lemma \ref{lemma6} gives the following theorem:

\begin{lemm}
If $a>a_c$ and $y>1$ then the free energy of a pulled adsorbing uniform
comb with $t$ teeth is given by
\[ \hspace{-2.5cm}
\zeta^{(t)}(a,y) = \lim_{m\to\infty} \sfrac{1}{Tm}\log K_{Tm}^{(t)}(a,y) =
\max\{\kappa(a), \sfrac{t+1}{2t+1}\lambda(y)+\sfrac{t}{2t+1}\log \mu_3, 
\sfrac{2t}{2t+1} \kappa(a) + \sfrac{1}{2t+1}\lambda(y)] \}. \quad\square
\]
\label{lemma7}
\end{lemm}

Next, consider the case $y\leq 1$.  

By monotonicity, and since $\lambda(1) = \log \mu_3$
and $\kappa(a) \geq \log \mu_3$,
\begin{equation}
\zeta^{(t)}(a,y) \leq \zeta^{(t)}(a,1) = \kappa(a) .
\label{eqn51}
\end{equation}
We now outline a proof for the corresponding lower bound.  An adsorbing
uniform comb can be constructed by concatenating in sequence 
$t+1$ doubly unfolded adsorbing loops to create a backbone.  
These loops are in the half-lattice $x_3\geq 0$ with both endpoints in the
adsorbing plane $x_3=0$, and they are doubly unfolded in the
$x_1$ and $x_2$ directions.  The projection of each unfolded loop
in the $x_1x_2$-plane falls within a rectangular region with endpoints
of the loop at opposite left-most bottom-most and right-most top-most
corners.  Thus, the projected backbone is contained in a union of a sequence
of rectangles joined at opposite corners, each containing the projection of
one loop.  The $t$ teeth are appended to vertices where the rectangles
join each other, and these vertices have height zero (that is, they are visits).  
Choose the teeth to be $\alpha$-bridges quarantined in disjoint $\alpha$-wedges 
which are also disjoint with the rectangles containing the projected backbone.  Since the 
limiting free energy of adsorbing doubly unfolded loops, and of $\alpha$-bridges,
is $\kappa(a)$, this gives the lower bound $\zeta^{(t)}(a,y) \geq \kappa(a)$.
With equation \Ref{eqn51} this gives the following lemma

\begin{lemm}
$\displaystyle
\zeta^{(t)}(a,y) = \kappa(a), \;\hbox{for all $y\leq 1$}. \quad\square$
\label{lemma8}
\end{lemm}

Finally, we consider the case $a\leq a_c$.  

By monotonicity, since $\kappa(a) = \log \mu_3$ if
$a\leq a_c$ and $\lambda(y) \geq \log \mu_3$,
\begin{equation}
\zeta^{(t)}(a,y) \leq \zeta^{(t)}(1,y) =
\sfrac{t+1}{2t+1}\lambda(y)+\sfrac{t}{2t+1}\log \mu_3 .
\label{eqn52}
\end{equation}
Next, $\zeta^{(t)}(0,y)$ is the free energy of pulled combs with zero
visits.  These are combs with a backbone from the origin, pulled at the other
endpoint, with first edge from $(0,0,0)$ to $(0,0,1)$, and with all remaining
vertices disjoint with the plane $x_3=0$.  If the first edge is removed and the
origin moved to $(0,0,1)$, then this is an almost uniform comb with backbone
from $(0,0,1)$, in the half-space $x_3\geq 1$, pulled at its other endpoint,
with $t$ teeth, and with all branches of the same length, except the branch
from $(0,0,1)$, which has length 1 edge shorter than the remaining branches.
If $y\leq 1$ then the free energy of this almost uniform comb is $\log \mu_3$,
by the methods in reference
\cite{Soteros}.  This shows that $\log \mu_3 \leq \zeta^{(t)}(a,y) \leq \log \mu_3$
if $y\leq 1$ and $a\leq a_c$, so that $\zeta^{(t)}(a,y) = \log \mu_3$ in this regime.
If $y>1$, then the arguments similar to those leading to lemma \ref{lemma7}, shows
that the limiting free energy of the almost uniform comb is given by
$\zeta^{(t)}(1,y)$.  That is, $\zeta^{(t)}(0,y)=\zeta^{(t)}(1,y)$, and so comparison to equation
\Ref{eqn52} then gives the following lemma.

\begin{lemm}
$\displaystyle
\zeta^{(t)}(a,y) = \sfrac{t+1}{2t+1}\lambda(y)+\sfrac{t}{2t+1}\log \mu_3, 
\;\hbox{for all $a\leq a_c$}. \quad\square$
\label{lemma9}
\end{lemm}

By comparing lemmas \ref{lemma7}, \ref{lemma8} and \ref{lemma9} the
free energy of pulled adsorbing uniform $t$-combs is given in theorem
\ref{thm10}.

\begin{theo}
The free energy of a pulled adsorbing uniform
comb with $t$ teeth is given by
\[ \hspace{-2.5cm}
\zeta^{(t)}(a,y) = \lim_{m\to\infty} \sfrac{1}{Tm}\log K_{Tm}^{(t)}(a,y) =
\max\{\kappa(a), \sfrac{t+1}{2t+1}\lambda(y)+\sfrac{t}{2t+1}\log \mu_3, 
\sfrac{2t}{2t+1} \kappa(a) + \sfrac{1}{2t+1}\lambda(y)  \}. \quad\square
\]
\label{thm10}
\end{theo}

\begin{figure}[h!]
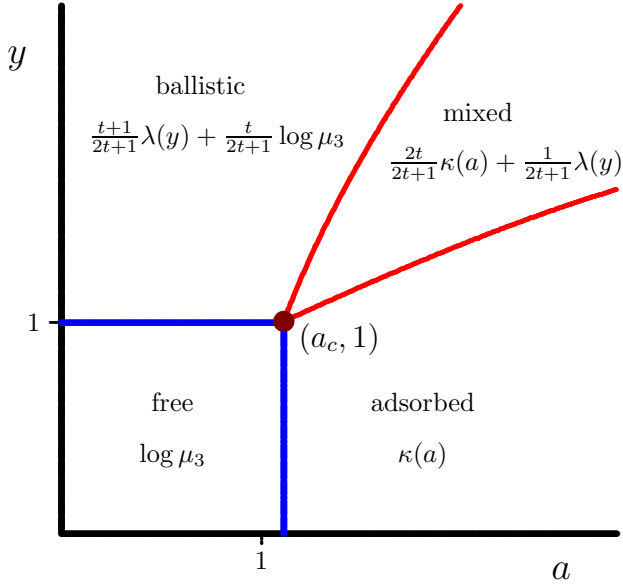

\beginpicture
\setcoordinatesystem units <2.1pt,2pt>
\setplotarea x from -40 to 100, y from -10 to 100

\color{black}
\setplotsymbol ({$\cdot$})
\plot -2 40 0 40 /   \plot 36 -2 36 0 /

\setsolid

\setplotsymbol ({\tiny$\bullet$})
\plot 0 100 0 0 100 0 /

\color{black}
\put {\Large$y$} at -8 90
\put {\Large$a$} at 90 -7
\put {$1$} at -5 40
\put {$1$} at 36 -5
\put {\large$(a_c,1)$} at 50 37

\put {\hbox{free}} at 20 25 
\put {\hbox{$\log \mu_3$}} at 20 15 

\put {\hbox{ballistic}} at 25 85 
\put {\hbox{$\sfrac{t+1}{2t+1}\lambda(y)+\sfrac{t}{2t+1}\log \mu_3$}} at 28 75 

\put {\hbox{adsorbed}} at 65 25 
\put {\hbox{$\kappa(a)$}} at 65 15

\put {\hbox{mixed}} at 75 80 
\put {\hbox{$\sfrac{2t}{2t+1} \kappa(a) + \sfrac{1}{2t+1}\lambda(y)$}} at 80 70

\color{blue}
\plot 40 0 40 40 0 40 /

\setplotsymbol ({\LARGE$\cdot$})
\color{red}
\setquadratic
\plot 40 40 53 69 72 100 /
\plot 40 40 73 55 100 65 /
\setlinear
\color{Maroon}
\put {\huge$\bullet$} at 40 40

\color{black}
\normalcolor
\endpicture
\caption{The phase diagram of pulled adsorbing $t$-combs.
The phase boundaries separating the adsorbed, mixed and ballistic phases
are first order. The phase boundary between the adsorbed and mixed
phases is given by the solution of $\lambda(y)=\kappa(a)$, and the phase
boundary between the mixed and ballistic phases by the solution of
$\lambda(y) + \log \mu_3 = 2\,\kappa(a)$.}
\label{figure22}
\end{figure}

There is a mixed phase in this model.  For $y > 1$ and $\sfrac{1}{2}(\lambda(y)+\log \mu_3)
< \kappa(a) < \lambda(y)$ the free energy is given by $\zeta^{(t)}(a,y)= \sfrac{1}{2t+1}
(2t\,\kappa(a)+\lambda(y))$.  So if the force is small (and $a$ is large and fixed so that
$\kappa(a)>\lambda(y)$),
the free energy is $\kappa(a)$ and the comb is adsorbed.  As $y$ increases  there
is a transition to the mixed phase where the force pulls  the last  backbone branch off the surface
but leaves the rest of the comb adsorbed.   In this mixed phase the
free energy is given by 
$\sfrac{2t}{2t+1} \kappa(a) + \sfrac{1}{2t+1}\lambda(y)$.
Further increasing  $y$, so that
$\kappa(a) <\sfrac{1}{2}(\lambda(y)+\log \mu_3)$, takes the model through a transition
to a phase with free energy $\zeta^{(t)}(a,y) = \sfrac{t+1}{2t+1}\lambda(y)+\sfrac{t}{2t+1}\log \mu_3$.
That is, the backbone (of length $(t+1)n$) is ballistic, and the teeth have all been
pulled from the adsorbing plane and contribute $\sfrac{t}{2t+1}\log \mu_3$ to the
free energy.

In other words, if the force is large enough to pull off the last tooth, then it will pull off
all the teeth.

\section{Discussion}
\label{sec:discussion}
The primary aim of this paper has been to extend previous work on self-avoiding
walk models of polymers, adsorbed at a surface and pulled off by application of a force,
as in an AFM experiment, for example.  Previous papers have focussed on linear
polymers (modelled by self-avoiding walks) \cite{Rensburg2013}, ring polymers (modelled by lattice
polygons) \cite{Guttmann2018}
and 3-star polymers \cite{Rensburg2018}.  We have extended this work to a variety of 
other branched polymer architectures, including stars with different 
numbers of arms, tadpoles, dumbbells, and combs.  All of these can be modelled 
by variants of self-avoiding walks and lattice polygons.  Our main result is that the
phase diagram of a pulled adsorbing branched polymer is dependent on its
homeomorphism type.  For example, the phase diagram of pulled adsorbing $f$-stars
depends on $f$, and also on the location where the star is pulled, and these phase
diagrams are different from the phase diagrams of pulled adsorbing tadpole, 
dumbbell or comb architectures.

In the case of linear and ring polymers (in three dimensions, on the simple
cubic lattice) it has previously been shown
that there are three phases, a free phase, an adsorbed phase and a ballistic
phase \cite{Guttmann2018,Rensburg2013}.  For 3-star polymers 
there is an additional mixed phase where
the free energy depends on both the magnitude of the force and the magnitude
of the surface interaction \cite{Rensburg2018}.  
We have shown that other homeomorphism types can
also exhibit four phases, but we have not found any cases with more
than four phases.  4-stars, 5-stars and 6-stars (pulled at a vertex of unit degree)
all have similar phase diagrams
to 3-stars.  In the case of tadpoles it depends on how the tadpole is attached
to the surface and on where the force is applied.  We have found examples 
with three and with four phases.  For combs we always find four phases, independent of
the number of teeth on the comb.

Our results considerably extend the number of polymer architectures
(or homeomorphism types) that have been studied and show how the 
phase diagram depends on the architecture.

\section*{Acknowledgement}
This research was partially supported by  Discovery Grants from NSERC of Canada.

\end{document}